\documentclass[ba]{imsart}

\RequirePackage{amsthm,amsmath,amsfonts,amssymb}
\RequirePackage[authoryear]{natbib}
\RequirePackage[colorlinks,citecolor=blue,urlcolor=blue]{hyperref}
\RequirePackage{graphicx}
\RequirePackage{multirow}
\usepackage{subcaption}
\usepackage{caption}
\usepackage{array}
\usepackage{adjustbox}
\usepackage{xcolor}
\usepackage{rotating}
\usepackage{lscape}
\usepackage{comment}
\usepackage{makecell}
\usepackage{bm}  
\startlocaldefs
\newcommand{\con}{{\,\vert \,}}

\newcommand{\half}{{\frac{1}{2}}}

\newcommand{\ton}{{\tau_{\omega,\nu}}}
\newcommand{\hg}{{\hat{\lambda}_g}}
\newcommand{\hf}{{\hat{\lambda}_f}}

\newcommand{\bt}{{\boldsymbol \theta}}

\newcommand{\bfo}{{\boldsymbol \omega}}

\pubyear{2024}
\arxiv{2010.00000}
\volume{TBA}
\issue{TBA}
\firstpage{1}
\lastpage{1}

\theoremstyle{plain}

\newtheorem{theorem}{Theorem}[section]
\newtheorem{lemma}[theorem]{Lemma}

\newtheorem{corollary}{Corollary}
\usepackage{amsmath}

\theoremstyle{remark}

\endlocaldefs
\newcommand{\DEP}{{\rm D_{EP}}( f\parallel g)}
\newcommand{\dep}{{\rm D_{EP}}}
\newcommand{\PED}{{\rm D_{EP}}( g\parallel f)}
\begin{document}

\begin{frontmatter}
\title{On Bayes factor functions}

\runtitle{On Bayes factor functions}

\begin{aug}
\author[A]{\fnms{Saptati} \snm{Datta}\ead[label=e2]{saptati@tamu.edu}},
\author[B]{\fnms{Riana} \snm{Guha}\ead[label=e3]{riana.guha@tamu.edu}},
\author[C]{\fnms{Rachael} \snm{Shudde}\ead[label=e4]{rachael.shudde@gmail.com}},
\author[D]{\fnms{Valen E.} \snm{Johnson}\ead[label=e1]{vejohnson@tamu.edu}}
\address[D]{Department of Statistics, Texas A\&M University\printead[presep={,\ }]{e1}}
\end{aug}

\begin{abstract}
Bayes Factor Functions (BFFs) represent Bayes factors as functions of prior hyperparameters \citep{Johnson2023, SPL2025}. While earlier work has primarily focused on normal-moment and gamma priors, we build on recent results by \citet{Wagenmakers2025} to explore broader classes of priors for constructing BFFs. For tests based on common test statistics, our findings show that inverse-moment and inverse-gamma priors can yield higher expected posterior probabilities in favor of both true null and true alternative hypotheses than other default priors.  Using test statistics from replicated experiments, we also demonstrate how method-of-moments estimators can inform prior parameter selection within these families, thereby enhancing inference in replicated studies and meta-analyses.
\end{abstract}



\end{frontmatter}


\section{Introduction}
Bayes factors provide a compelling alternative to $p$-values for summarizing hypothesis test outcomes. Unlike classical significance testing, they quantify evidence for both null and alternative hypotheses, yielding an interpretable measure of how plausible the observed data are under competing models. Introduced by Jeffreys in 1935, Bayes factors are defined as ratios of marginal likelihoods that directly measure evidential weight in hypothesis testing \citep{Jeffreys1935, Jeffreys1939}\footnote{Haldane (1932) had already introduced a mixture prior formulation that anticipated similar ideas \citep{Haldane1932, EtzWagenmakers2017}}. Building on Jeffreys’ foundation, \citet{Berger1985} established their role in Bayesian model choice and clarified their link to posterior model probabilities. Later, \citet{Kass1995} described computational strategies and refined interpretive guidelines, further encouraging their application.

In the last two decades, many have argued that Bayes factors should replace or supplement $p$-values as the primary measure of evidence against a null hypothesis \citep[e.g.,][]{Bayarri2004, Johnson2013, Held2018_annrev, Rougier2019, Benjamin2017, Benjamin2019ThreeRecommendations}. Nonetheless, their adoption has been limited by the need to specify both prior densities on model parameters and prior probabilities for the hypotheses themselves. When only two hypotheses are compared, the latter issue is often avoided by assigning equal prior weights or simply reporting the Bayes factor. However, defining suitable alternative priors remains a central challenge, and numerous proposals have been developed.

Jeffreys \citeyearpar{Jeffreys1939} was the first to suggest a Cauchy prior for testing whether a normal mean equals zero, laying the groundwork for the Jeffreys–Zellner–Siow (JZS) prior \citep{Zellner1980, Zellner84}. The JZS prior, based on a scaled Cauchy distribution, was later operationalized by \citet{Rouder2009} and incorporated in the \texttt{BayesFactor} package by \citet{Morey2023}. Other important contributions include intrinsic priors \citep{Berger1996, Berger1996, Moreno1998, MorenoBertolinoRacugno1998, CasellaMoreno2006, BergerPericchi2001}, which require specification of a minimal training sample size, and fractional Bayes factors \citep{OHagan1995, OHagan1997}, which depend on selecting a training fraction of the data. Conjugate $g$-priors \citep[e.g.,][]{Zellner84, Gprior_Zellner, Liang2008} offer another approach, later extended to generalized linear models by \citet{Held2011}. \cite{Bayarri2012} and \cite{Bayarri2012_Conventional} provide criteria that objective priors should satisfy and show that many previously proposed priors, with modification, do. 
Nonlocal priors represent yet another class. They vanish at parameter values consistent with the null hypothesis and yield faster accumulation of evidence in favor of both true null and true alternative hypotheses \citep{Johnson2010}, but they also require specification of scale parameters. A comprehensive overview of these and other objective priors is provided by \citet{Consonni2018}.

A persistent difficulty across these approaches is the selection of hyperparameters that determine prior scale. To address this problem, we advocate reporting Bayes factor functions (BFFs) \citep{Johnson2023}. Rather than presenting a single Bayes factor, BFFs summarize the continuum of Bayes factors obtained from a class of alternative priors.

In this article, we focus on BFFs defined using the sampling distributions of test statistics rather than full data likelihoods \citep{Johnson2005}. This choice simplifies matters in several ways. First, the null distributions of classical test statistics are known, and their alternatives often depend only on a scalar noncentrality parameter, which simplifies prior specification. Second, most hypothesis tests reported in the social, biological, and medical sciences are based on classical test statistics, making this approach widely applicable \citep[e.g.][]{Skaik2015,LiuWang2021}. Third, constructing BFFs from test statistics facilitates the aggregation of evidence across studies, enabling systematic reviews and meta-analyses.

Of course, BFFs can also be defined using full likelihoods and multidimensional parameter priors. In such cases, Bayes factor surfaces can be used to display variation as a function of prior hyperparameters \citep{Franck2020}. However, these are often more difficult to interpret in routine testing settings and less convenient for aggregating replicated results. A critical feature of our approach is that Bayes factors are expressed as interpretable functions of the discrepancy between null and alternative hypotheses.

The remainder of this paper proceeds as follows:
\begin{itemize}
\item Section 2 formally defines BFFs, illustrating their construction with inverse-moment priors for tests of normal means, and compares them informally to BFFs based on Cauchy priors \citep{Jeffreys1961, Zellner1980}.

\item Section 3 applies an equality from \citet{Wagenmakers2025} to compare expected posterior probabilities across different prior classes, including inverse-moment, $g$-prior, Cauchy, normal, gamma, and inverse-gamma densities.

\item Section 4 addresses the choice of shape and scale parameters for inverse-moment and inverse-gamma priors and proposes a procedure for indexing BFFs by standardized effect sizes.

\item Section 5 presents worked examples demonstrating the construction of BFFs.

\item Section 6 analyzes replicated experiments, showing how nonlocal priors induce a natural hierarchical structure on effect sizes across studies by scaling noncentrality parameters according to sample size, thereby allowing them to be modeled as draws from a common prior on effect size.
\end{itemize}

\section{Construction of BFFs}\label{bffconst}

To illustrate the construction of a BFF based on a test statistic $x$, we assume the following parametric Bayesian framework inspired by null hypothesis significance testing (see \cite{LyWagenmakersRousseau2007} and comments in \cite{Petrone} for connections to interval null hypothesis tests). Specifically, we define the hypotheses as follows: 
\begin{eqnarray*}
H_0 &:& x\sim f(\cdot \mid \lambda_0), \qquad \lambda_0  \ \mbox{is known},\\
H_1 &:& x \sim f(\cdot \mid \lambda), \qquad \lambda \sim g(\cdot\mid \psi).\\
\end{eqnarray*}
Here, $f(\cdot\mid\lambda)$ denotes the distribution of the test statistic parameterized by a non-centrality parameter $\lambda$, and $g(\cdot\mid\ \psi)$ is a parametric family of prior distributions on $\lambda$ under the alternative hypothesis $H_1$, indexed by $\psi$.  
Given a fixed value of $\psi$, the Bayes factor for the test can be expressed
\[
BF_{10}(x\con \psi) = \frac{\int_\Lambda f(x \mid \lambda)g(\lambda \mid \psi)d\lambda}{f(x \mid \lambda_0)},
\]
where $\Lambda$ denotes the parameter space of $\Lambda$.

The Bayes factor function for a fixed value of the test statistic 
$x$ is defined as a function of the hyperparameter $\psi$
taking values in a parameter space 
$\Psi$,
\[ 
\mbox{BFF}_x : \Psi \rightarrow [0, \infty), \quad \psi \mapsto BF_{10}(x \mid \psi),
\]
where $BF_{10}(x\con\psi)$ is the Bayes factor for the observed value of $x$ under the specification indexed by $\psi$.
When the dependence on $x$ is understood from context, we refer to this function simply as the BFF.

From the above discussion, if $\psi \mapsto \xi$ is a bijection onto its image, it follows that the BFF can be equivalently reparameterized as a function of
$\xi$. That is, there exists a function
\[
\mbox{BFF}_x^\xi : \Xi \rightarrow [0, \infty), \quad \xi \mapsto BF_{10}(x \mid \psi(\xi)),
\]
where $\Xi$ denotes the image of the bijection 
$\psi \mapsto \xi$, and $\psi(\xi)$ is the inverse mapping defining $\psi$ as a function of $\xi$.  A key component of the methodology advocated here is determining parameterizations of $\psi$ that yield scientifically interpretable BFFs.  


To illustrate the construction of a BFF, consider a two-sided $z$-test with data $x_1,\dots,x_n$ independently distributed as $N(\mu,\sigma^2)$, where $\sigma^2$ is known. The standardized test statistic is $z = \sqrt{n}x/\sigma$, which follows a $N(\sqrt{n}\omega,1)$ distribution, with $\omega := \mu/\sigma$ denoting the standardized effect. The null hypothesis specifies $\mu = 0$, or equivalently $\omega = 0$. For the alternative, rather than placing a prior directly on $\mu$, we specify a prior on the noncentrality parameter $\lambda := \sqrt{n}\omega$.

\begin{figure}[ht!]
    \centering
    \includegraphics[width=12cm]{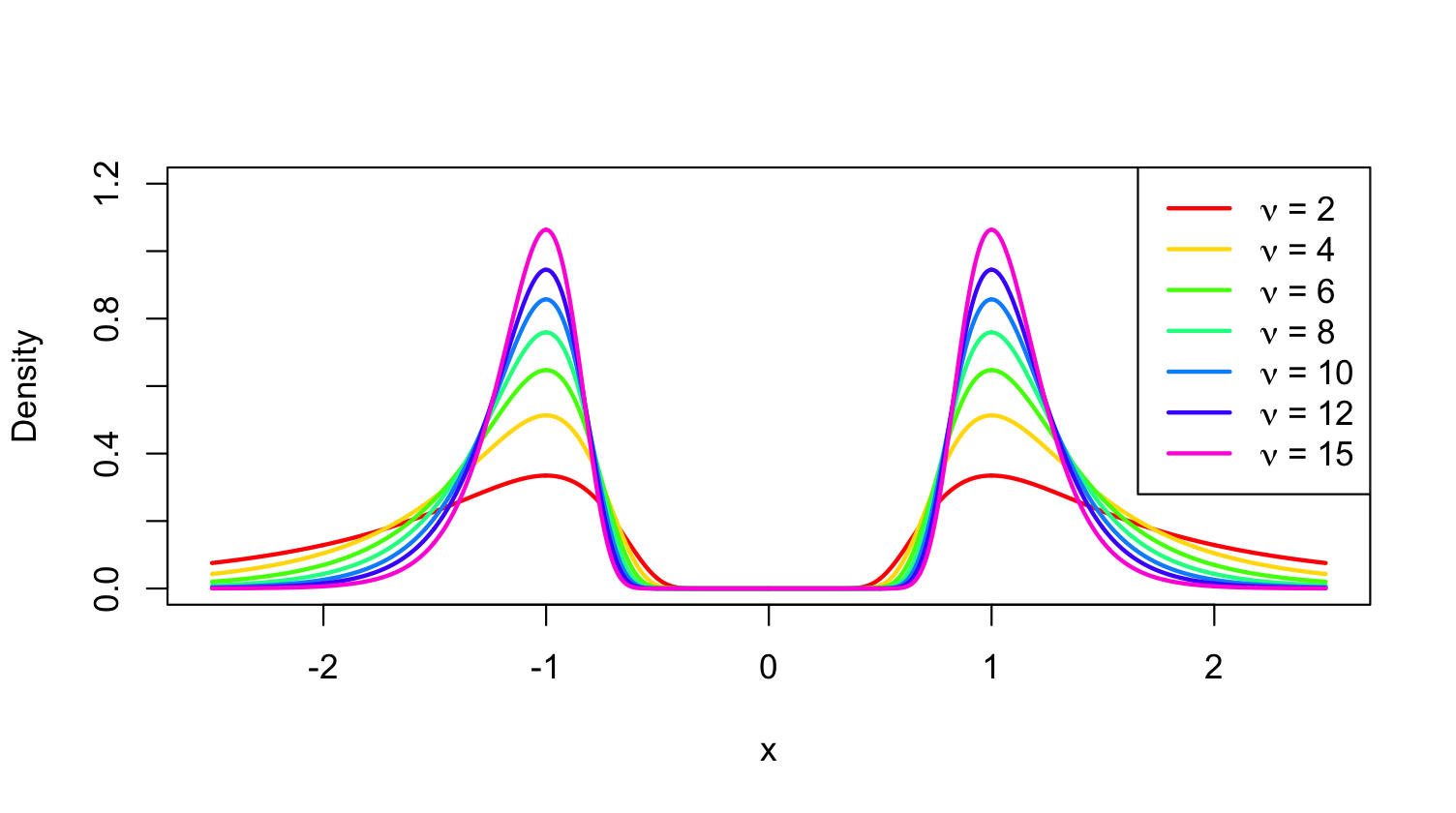}
    \caption{Inverse moment priors with modes at \(\pm 1\) and varying values of \(\nu\).}
    \label{imdensities}
\end{figure}

As a default prior for $\lambda$, we adopt the inverse-moment distribution \citep{Johnson2010}, defined by
\begin{equation}\label{eqn:prior_Density_z}
i(\lambda \mid \tau, \nu) = \frac{\tau^{\nu/2}}{\Gamma(\nu/2)} (\lambda^2)^{-(\nu+1)/2} \exp\left(-\frac{\tau}{\lambda^2}\right), \quad -\infty < \lambda < \infty,\ \tau,\nu > 0.
\end{equation}
This density has modes at
\begin{equation}\label{invmode}
\lambda = \pm \sqrt{2\tau/(\nu+1)}.
\end{equation}
For one-sided alternatives, the densities $i^+(\cdot\mid \tau,\nu)$ and $i^-(\cdot\mid \tau,\nu)$ are defined as \newline $2i(\cdot\mid \tau,\nu)$, restricted to the positive or negative half-line, respectively. Several examples of these priors are shown in Fig.~\ref{imdensities}. We denote the corresponding distribution by $\mathcal{I}(\lambda \mid \tau, \nu)$.

With $(\tau,\nu)$ fixed, the hypotheses can be written as
\[
H_0: z \sim N(0,1),
\]
\[
H_1: z \mid \lambda \sim N(\lambda,1), \quad \lambda \mid \tau,\nu \sim \mathcal{I}(\lambda\con \tau, \nu).
\]
Under $H_0$, the marginal density of $z$ is standard normal. Under $H_1$, the marginal density is
\[
m_1(z \mid \tau, \nu) = \int_{-\infty}^{\infty} \frac{1}{\sqrt{2\pi}} \exp\left[-\frac{(z-\lambda)^2}{2}\right] \frac{\tau^{\frac{\nu}{2}}}{\Gamma\left(\frac{\nu}{2}\right)} (\lambda^2)^{-\frac{\nu + 1}{2}} \exp\left(-\frac{\tau}{\lambda^2}\right) \, d\lambda.
\]
Although this integral lacks a closed form, it can be readily computed numerically in software such as R,
and its logarithm, $\log BF_{10}(z \mid \tau, \nu)$, is called the weight of evidence (WOE) \citep{Good1985,Kass1995}.

Different values of $(\tau, \nu)$ correspond to different alternative hypotheses.  In defining the BFF for this test, it is helpful to specify $(\tau, \nu)$ based on hypothesized prior modes under the alternative hypothesis. For instance, suppose that a hypothesized prior mode for the noncentrality parameter under the alternative hypothesis is $\lambda_1 = 
\sqrt{ n}\omega_1$. This noncentrality parameter corresponds to the hypothetical standardized effect $\omega_1$.  Assuming that $\nu$ is fixed, then from equation (\ref{invmode}) the value of $\tau$ that places the prior mode at $\lambda_1$ is
\[ \tau_{\omega_1, \nu} = \frac{n(\nu + 1)\omega_1^2}{2}. \]
Thus, the Bayes factor can be expressed as a function of the standardized effect $\omega_1$, and a BFF can be defined as the mapping of standardized effects to Bayes factors obtained as $\omega_1$ is varied over its domain.
Criteria for selecting $\nu$ for this particular class of priors are discussed in Section 4.

\begin{figure}[ht!]
    \centering
    \includegraphics[width=4.5in]{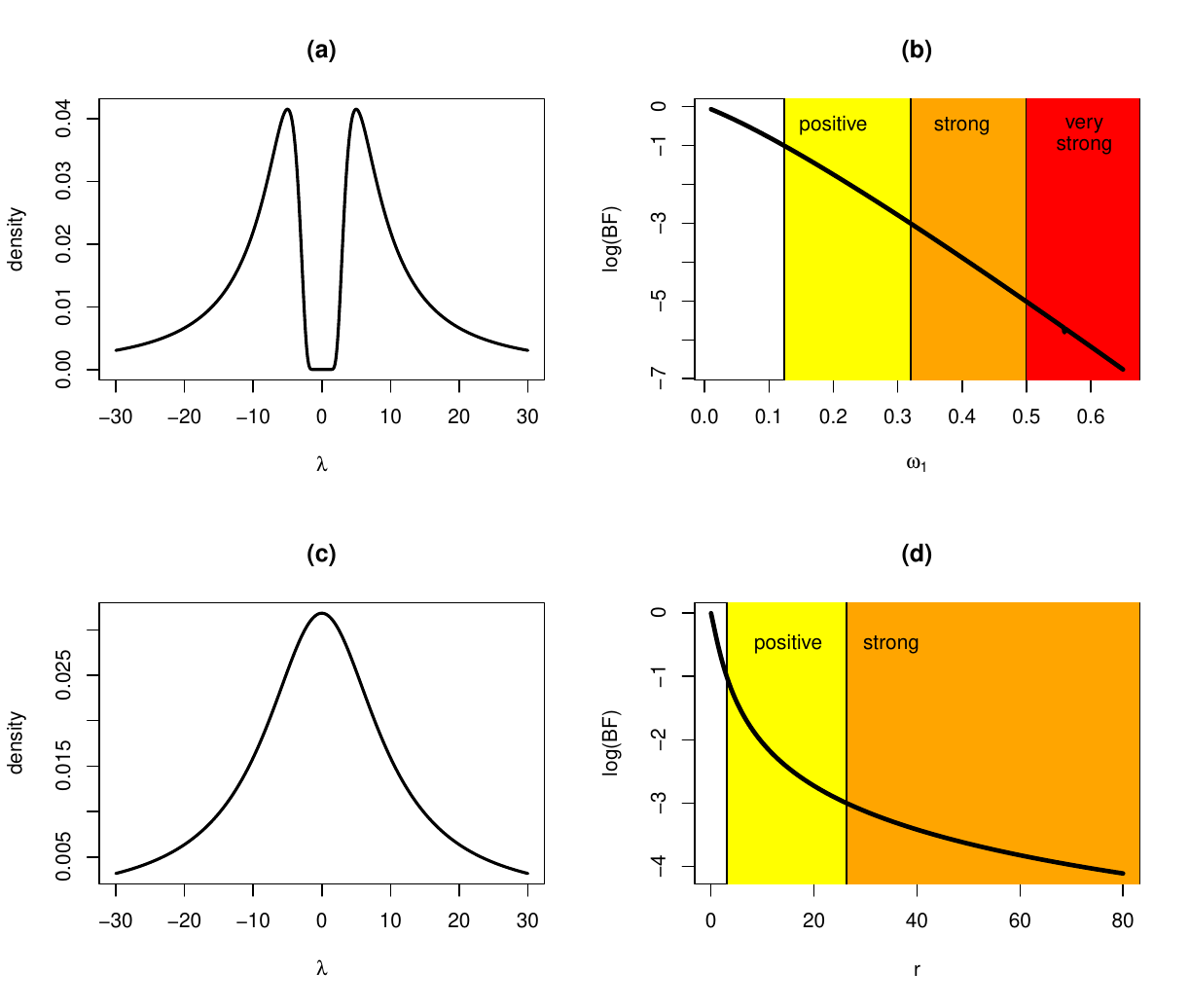}
    \caption{Prior densities and BFFs. (a) Inverse moment prior density for \(\nu = 9\), \(\omega_1 = 0.5\), $n=100$. This density has its modes at $\lambda = \pm \sqrt{n}\omega_1 = \pm 5$. (b) BFF generated from inverse moment prior density, on logarithmic scale. This plot displays strong evidence (i.e., $\log[BF_{10}(z)]<-3$) against alternative prior densities corresponding to $\omega_1>0.32$, and very strong evidence ($<-5$) for $\omega_1>0.50$. (c) Cauchy prior on noncentrality parameter with scale $r=5$. (d) BFF based on Cauchy prior as scale parameter $r$ is increased.} 
    \label{zprior}
\end{figure}

To make this example more concrete, suppose $z=1.0$ based on a sample size of $n=100$.  Fixing $\nu=1$, an alternative prior density corresponding to $\omega_1 = 0.5$ is depicted in Fig.~\ref{zprior}(a).  This curve represents the inverse-moment prior on $\lambda$ with $\nu=1$ and $\tau_{0.5,1}=100(2)(0.5^2)/2 = 25$.  The BFF obtained using this prior and varying $\omega_1$ is displayed in Fig.~2(b).  The Bayes factor corresponding to the prior in Fig.~2(a) is located at the boundary between the orange- and red-shaded regions. 

Several features of Fig.~\ref{zprior}(b) deserve mention.  With $z=1$, the BFF is a decreasing function of $\omega_1$. The BFF demonstrates positive support for the null hypothesis for inverse-moment priors centered on $\omega_1> 0.12$ ($\mbox{WOE}<-1$), strong support for $\omega_1>0.32$ ($\mbox{WOE}<-3$), and very strong support for $\omega_1>0.50$ ($\mbox{WOE}<-5$).  
This figure also shows that the log(BFF) converges to 0 as $\omega_1$ goes to 0.  This property holds for all values of the $z$ statistic because the inverse-moment prior densities centered on small $\omega$ values concentrate their mass near the null parameter value, resulting in approximately the same marginal probability being assigned to the data under both hypotheses.  

The same construction extends to $t$, $\chi^2$, and $F$ tests. For $\chi^2_k$ and $F_{k,m}$ statistics, inverse-gamma priors $IG(\nu/2,\tau)$ on the noncentrality parameter provide a natural analogue, as $W \sim \mathcal{I}(\nu,\tau)$ implies $W^2 \sim IG(\nu/2,\tau)$. Under these priors, the Bayes factor for a two-sided $z$ statistic equals that for the $\chi^2_1$ statistic $z^2$, and likewise for a two-sided $t_k$ and its corresponding $F_{1,k}$ statistic (provided the same values of $\nu$ and $\tau$ are used to define the tests).



While inverse-moment and inverse-gamma priors serve as convenient defaults, BFFs can also be defined using other prior distributions for the non-centrality parameters. For example,
 Fig~\ref{zprior}(c) and (d) illustrate the definition of a BFF using a Cauchy prior for the $z$-test described above \citep[e.g.,][]{Rouder2009,Rouder2012}.  The scale parameter $r$ of the Cauchy prior in Fig.~\ref{zprior}(c) is five, and its interquartile range 10, equaling the distance between the modes of the inverse-moment prior in Fig.~\ref{zprior}(a). The tail behaviors of the two prior densities are similar. The BFF based on the Cauchy prior depicted in Fig.~\ref{zprior}(d) provides ``positive'' evidence in favor of the null hypothesis for priors with $r>3$, and ``strong evidence'' for $r>26$.  

Direct visual comparison of the BFF curves in Fig.~2 is complicated by the fact that they are indexed by different parameters (i.e., $\omega_1$ and $r$). To address this issue, and to enable broader comparisons of BFFs constructed under alternative default priors, the following section introduces a methodology for evaluating and comparing the impact of prior specification on test outcomes based on BFFs.

\section{Comparisons of alternative classes of prior distributions}

The asymptotic behavior of Bayes factors under true point null hypotheses is largely determined by the properties of the alternative prior density at the null value of the parameter being tested \citep{Johnson2010}. In the present framework, hypothesis tests assess whether the noncentrality parameter $\lambda$ of a test statistic equals zero. Consequently, the behavior of the prior distribution for $\lambda$ near zero directly affects the properties of the Bayes factor.

When the alternative prior on $\lambda$ is continuous and strictly positive at zero—corre-\\sponding to a continuous local alternative—and under certain regularity conditions, the Bayes factor in favor of the alternative hypothesis, when the null is true, converges at the rate $BF_{10} = O_p(n^{-1/2})$ \citep{Johnson2010}. For specific test statistics, such as the $z$, $t$, $\chi^2$, and $F$ statistics, \citet{SPL2025} derived closed-form expressions for Bayes factors based on normal moment priors of the form
\begin{equation} \label{normMom}
p(\lambda) = \frac{(\lambda^2)^r}{(2\tau^2)^{r+\half} \Gamma(r+\half)} \exp\left(-\frac{\lambda^2}{2\tau^2}\right).
\end{equation}
imposed on noncentrality parameters $\lambda$ of $z$ and $t$ tests, and $Gamma(\lambda\con k/2+r,1/(2\tau^2)$ priors for $\chi^2_k$ and $F_{k,m}$ tests.
Their results show that Bayes factors in favor of a true null hypothesis satisfies $BF_{10} = O_p(n^{-r - 1/2})$ for $z$ and $t$ tests, and $BF_{10} = O_p(n^{-r - k/2})$ for $\chi^2_k$ and $F_{k,m}$ tests.

In contrast, if inverse-moment priors are assumed for the non-centrality parameters of $z$ and $t$ tests, and inverse gamma (IG) priors are imposed on noncentrality parameters for $\chi^2$ and $F$ tests, the convergence of resulting Bayes factors in favor of true null hypotheses is of exponential order in $\sqrt{n}$.  Specifically, the following theorems apply.
\begin{theorem}[{\em z} test]
    Suppose $x \sim N(\lambda,1)$, $H_0: \lambda=0$, and $H_1: \lambda\sim I(\lambda\con \tau,\nu)$, $\tau = \delta n$, $\delta, \nu>0$.  If $H_0$ is true and $\lambda = a \sqrt{n}$ for $a>0$ and increasing $n$, then
    \begin{equation*}
        BF_{10}(x) = O_p(\exp\left[-c\sqrt{n}\right]) \quad \mbox{for some} \quad 0<c<\half. 
    \end{equation*}
\end{theorem}
\begin{theorem}[{\em t} test]
    Suppose $x \sim T_k(\lambda)$, $k = bn, b >0$, $H_0: \lambda=0$, and $H_1: \lambda\sim I(\lambda\con \tau,\nu)$, $\tau = \delta n$, 
     $\delta, \nu>0$.  If $H_0$ is true and $\lambda = a \sqrt{n}$ for $a>0$ and increasing $n$, then
    \begin{equation*}
        BF_{10}(x) = O_p(\exp\left[-c\sqrt{n}\right]) \quad \mbox{for some} \quad 0<c<\half. 
    \end{equation*}
\end{theorem}
\begin{theorem}[$\chi^2$ test]
    Suppose $x \sim \chi^2_k(\lambda)$, $k \in \mathbb{N}$ is fixed, $H_0: \lambda=0$, and $H_1: \lambda\sim IG\left(\lambda\con \tau,\frac{\nu}{2}\right)$, $\tau = \delta n$, $\delta, \nu>0$.  If $H_0$ is true and $\lambda = a n$ for $a>0$ and increasing $n$,  then
    \begin{equation*}
        BF_{10}(x) = O_p(\exp\left[-c\sqrt{n}\right]) \quad \mbox{for some} \quad 0<c<\half. 
    \end{equation*}
\end{theorem}
\begin{theorem}[$F$ test]
    Suppose $x \sim F_{m_1,m_2}(\lambda)$, $m_2 =b n, b>0$, $m_1 \in \mathbb{N}$ is fixed, $H_0: \lambda=0$, and $H_1: \lambda\sim IG\left(\lambda\con \tau,\frac{\nu}{2}\right)$, $\tau = \delta n$, $\delta, \nu>0$.  If $H_0$ is true and $\lambda = a n$ for $a>0$ and increasing $n$,  then
    \begin{equation*}
        BF_{10}(x) = O_p(\exp\left[-c\sqrt{n}\right]) \quad \mbox{for some} \quad 0<c<\half. 
    \end{equation*}
\end{theorem}
\noindent Proofs of Theorems 3.1-3.4 are provided in the Supplemental Materials.

For true alternative hypotheses, $BF_{01}$ is typically $O_p(\exp(-cn))$ for some $c>0$ for all alternative priors that are continuous and positive at the (true) non-null parameter \citep{Bahadur}.  
In the case of finite samples, of course, asymptotic convergence rates can be misleading, and practical performance may differ substantially from  asymptotic expectations.

To evaluate the finite sample performance of Bayes factors, we draw on a recent result of Wagenmakers and Grasman, who exposed the following equality. Here, $f$ and $g$ are prior predictive distributions for two models, ${\cal{M}}_{f}$ and ${\cal{M}}_{g}$, each assigned equal prior probability. Then \cite{Wagenmakers2025} define the discrepancy measure
\begin{equation*}
   \DEP= \int f(x) \frac{f(x)}{f(x)+g(x)} dx = \int g(x) \frac{g(x)}{f(x)+g(x)} dx ,
\end{equation*}
with $\DEP =  \PED$.
When equal prior probability is assigned to the two models, this equality states that the expected posterior probability of ${\cal{M}}_f$ when ${\cal{M}}_f$ is true equals the expected posterior probability of ${\cal{M}}_g$ when ${\cal{M}}_g$ is true.  We write $\dep$ as shorthand for $\DEP$ when the compared models are clear from context. 

We can use this measure to compare BFFs defined by different alternative prior distributions.  Specifically, letting ${\cal M}_0$ denote the null model $H_0$, and ${\cal M}_{1a}$ and ${\cal M}_{1b}$ denote two alternative models, we compare ${\rm D_{EP}}(0\parallel 1a)$ to ${\rm D_{EP}}(0\parallel 1b)$. If ${\rm D_{EP}}(0\parallel 1a)>{\rm D_{EP}}(0\parallel 1b)$, then the expected posterior probability of the true model when the null is compared to ${\cal M}_{1a}$ is higher than the expected posterior probability of the true model when the null is compared to ${\cal M}_{1b}$, assuming equal prior probabilities are assigned to each model in both comparisons.   

To compare the efficiency of non-local priors in accumulating evidence against a null hypothesis, prior modes provide a natural indexing parameter. 
However, local alternative priors used in standard Bayesian null hypothesis significance tests are often centered on the null value, making the mode unsuitable for indexing. Additionally, certain priors--such as the Cauchy--do not possess finite moments and so cannot be indexed by means or standard deviations. To accommodate these cases, alternative priors can be indexed by their interquartile range (IQR) in two-sided tests and by their medians in one-sided tests. This parameterization provides an interpretable basis for comparing the performance of BFFs across a range of prior distributions.

To compare choices of alternative prior densities using this methodology, we begin by considering BFFs based on $t$-tests. The null hypothesis is that the non-centrality parameter of the test statistic satisfies $\lambda=0$.  The following alternative classes of prior densities on the non-centrality parameter are compared:
\begin{enumerate}
    \item Inverse-moment priors $i(\lambda\con \tau, \nu)$ with shape parameter $\nu=1$.
    \item Cauchy priors $C(\lambda \con 0, r)$, where $r$ is the scale parameter \citep[e.g.,][]{Jeffreys1961, Rouder2009}.  
    \item Normal priors, $N(\lambda\con 0,g)$ \citep[e.g.,][]{Zellner1980,Zellner84}.
    \item Normal-moment priors \citep{Johnson2023}, with $r=1$ in equation (\ref{normMom}).
\end{enumerate}
When a $t$ test arises from a test of whether a normal mean is zero based on a random sample $\{X_i\} \  \mbox{i.i.d.}\ N(\mu,\sigma^2)$, $i=1,\dots,n$, then $\lambda = \sqrt{n}\mu/\sigma $. The Cauchy prior (2) then corresponds to the Jeffreys–Zellner–Siow (JZS) prior, specified as
\[ \mu/\sigma \con r, \sigma^2_b \sim N(0,r^2\sigma^2_b), \qquad \sigma^2_b \sim \chi^{-2}_1, \qquad \sigma^2 \propto 1/\sigma^2,\]
where $\chi^{-2}_1$ denotes an inverse chi-squared distribution on one degree of freedom \citep{Rouder2009,Rouder2012}. Alternative (3) corresponds to a g-prior, also with a non-informative prior imposed on $\sigma^2$ \citep{Liang2008}. Alternative (4) is a special case of the class of normal moment priors described in \cite{SPL2025}. 

\begin{figure}[ht!]
    \centering
    \includegraphics[width=2.5in]{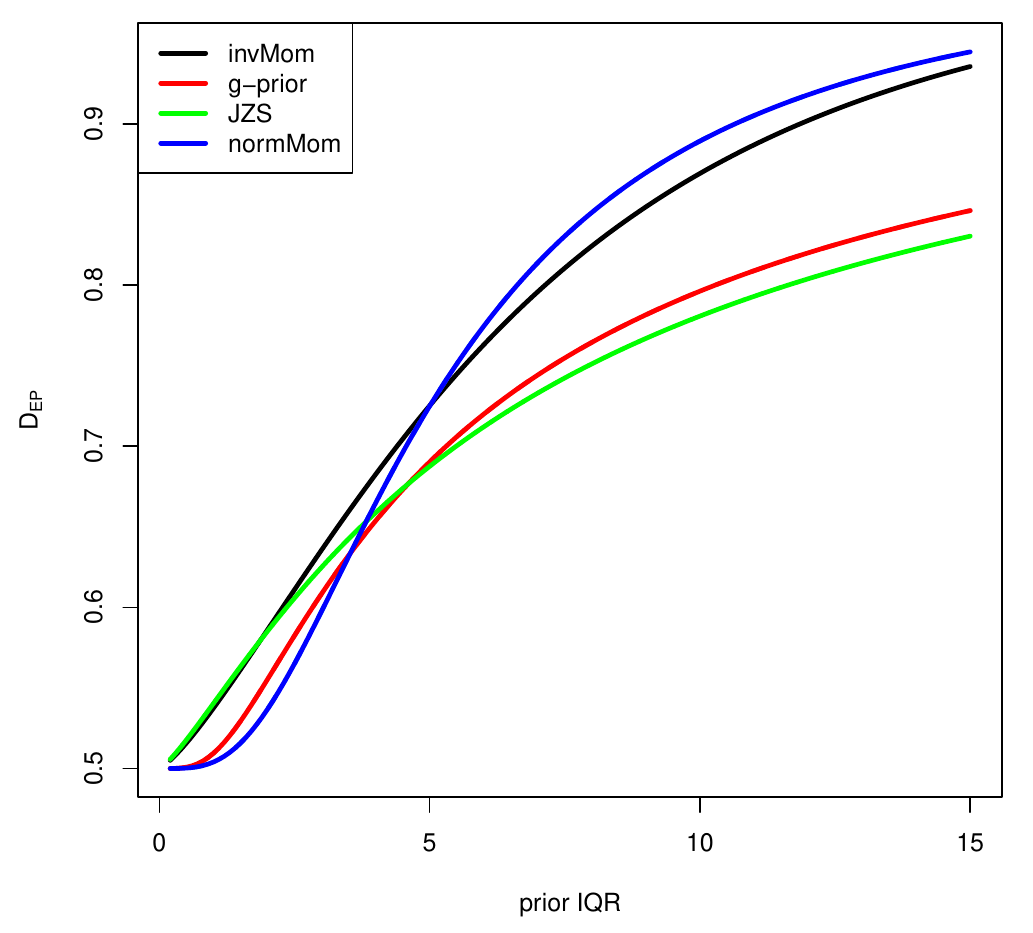}
    \caption{$\dep$ plots for $t$-test statistics with 15 degrees-of-freedom. The horizontal axis provides the IQR that each prior density assigns around the null value of the non-centrality parameter (i.e., $\lambda=0$). }. 
    \label{eppT}
\end{figure}

$\dep$ plots for these alternative prior densities versus the null hypothesis appear in Fig.~\ref{eppT}.  $\dep$ values are indexed by the IQR of each alternative prior density.  For example, the IQR of the Cauchy density is $2r$, while the IQR of the normal prior is $1.35\sqrt{g}$. The values of $\tau$ that correspond to a specified IQR of the inverse-moment and normal-moment priors were determined numerically.
The values in Fig.~3 reflect the $\dep$ associated with each alternative prior versus the null as its scale parameter is determined so that it assigns 0.5 probability to the same interval around the null value. 

Fig.~\ref{eppT} highlights several contrasts between the $\dep$ associated with the four alternative prior densities. When the IQR of the priors exceeds about 5.0, the non-local prior densities provide significantly higher $\dep$ than the local priors.  Recalling that the noncentrality of the $t$ density is $\lambda = \sqrt{n}\mu/\sigma$, an IQR of 5.0 corresponds to, for example, a sample size of 100 and a standardized effect of 0.5 (i.e., a medium effect size \citep{Cohen1988}). 

For small values of the prior IQR, the Cauchy and inverse-moment priors provide slightly better $\dep$, but, on average, all priors provide unconvincing evidence for the true model. Perhaps surprisingly, the JZS prior and the inverse-moment priors provide approximately the same $\dep$ for small prior IQRs.

\begin{figure}[ht!]
    \centering
    \includegraphics[width=4.5in]{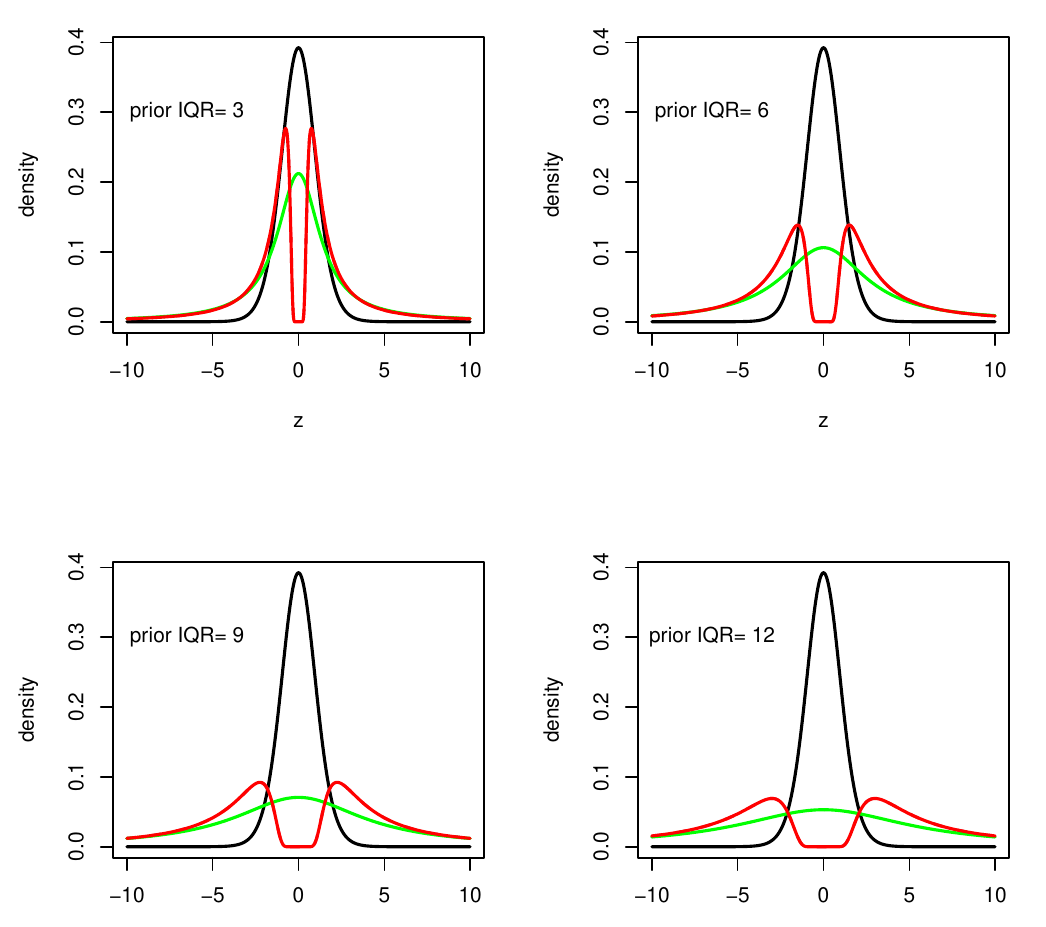}
    \caption{Plots of inverse-moment (red) and Cauchy (green) densities along with the null distribution of a $t_{15}$ test statistic (black). The IQRs of the inverse-moment and Cauchy densities range from 3 (upper right) to 12 (lower left).  } 
    \label{explanation}
\end{figure}

An explanation for the latter phenomenon can be gleaned from Fig.~\ref{explanation}.  This plot depicts Cauchy and inverse-moment alternative priors with varying IQRs along with the null distribution of a $t_{15}(0)$ statistic. When the prior IQR is 3, both priors concentrate much of their mass at values consistent with the sampling distribution under the null hypothesis.  Thus, the marginal density of data generated from the two alternative priors is approximately the same.

As the IQR of the prior densities increase, the inverse-moment prior allocates less mass near the origin at a faster rate than the Cauchy prior. This results in a more rapid increase in the ${\rm D_{EP}}$ statistic for the inverse-moment prior, as shown in Fig.~\ref{explanation}. This pattern persists with further increases in IQR, with the inverse-moment prior offering progressively greater support for the true hypothesis.


\begin{figure}[ht!]
    \centering
    \includegraphics[width=2.5in]{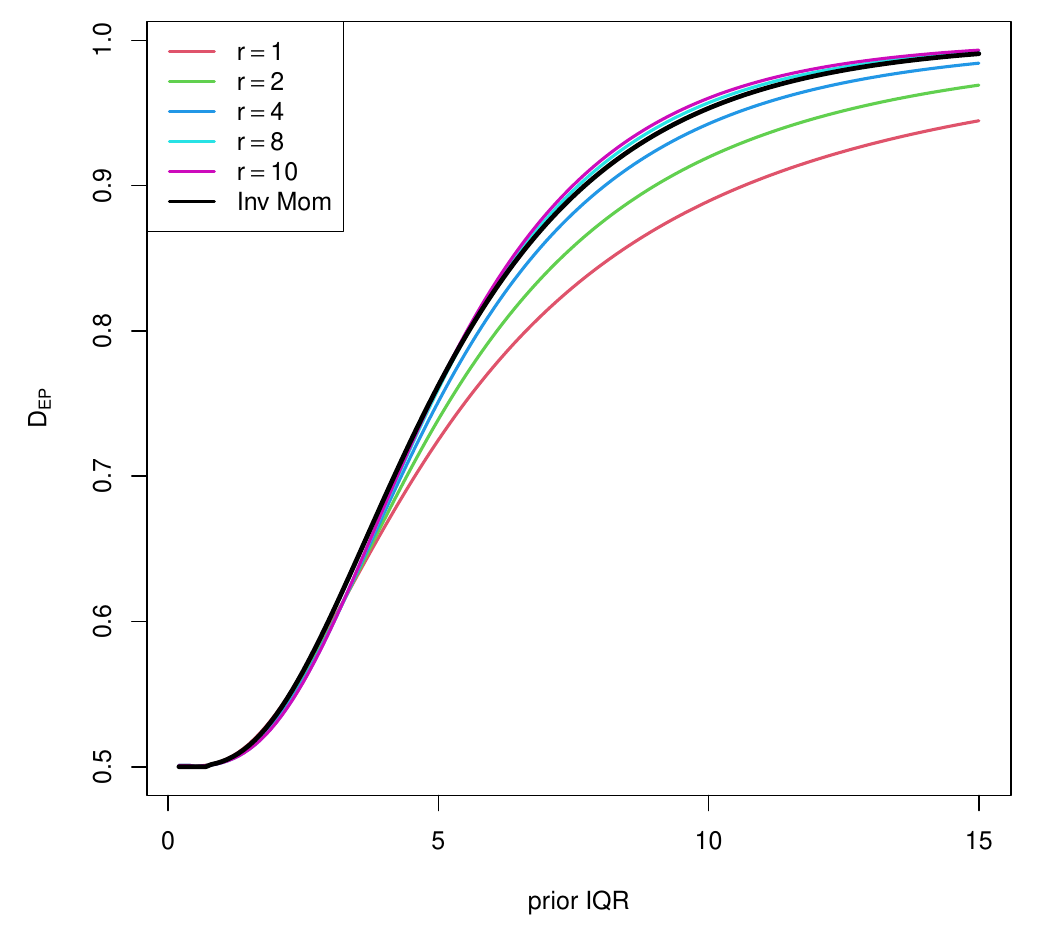}
    \caption{$\dep$ plots of normal moment priors of different orders, $r$, for testing whether the noncentrality parameter of a $t_{15}$ statistic is 0. For comparison, the $\dep$ plot for inverse moment alternative priors with shape parameter $\nu=9$ is provided by the black curve.} 
    \label{invMomvMom}
\end{figure}

\begin{figure}[ht!]
    \centering
    \includegraphics[width=2.5in]{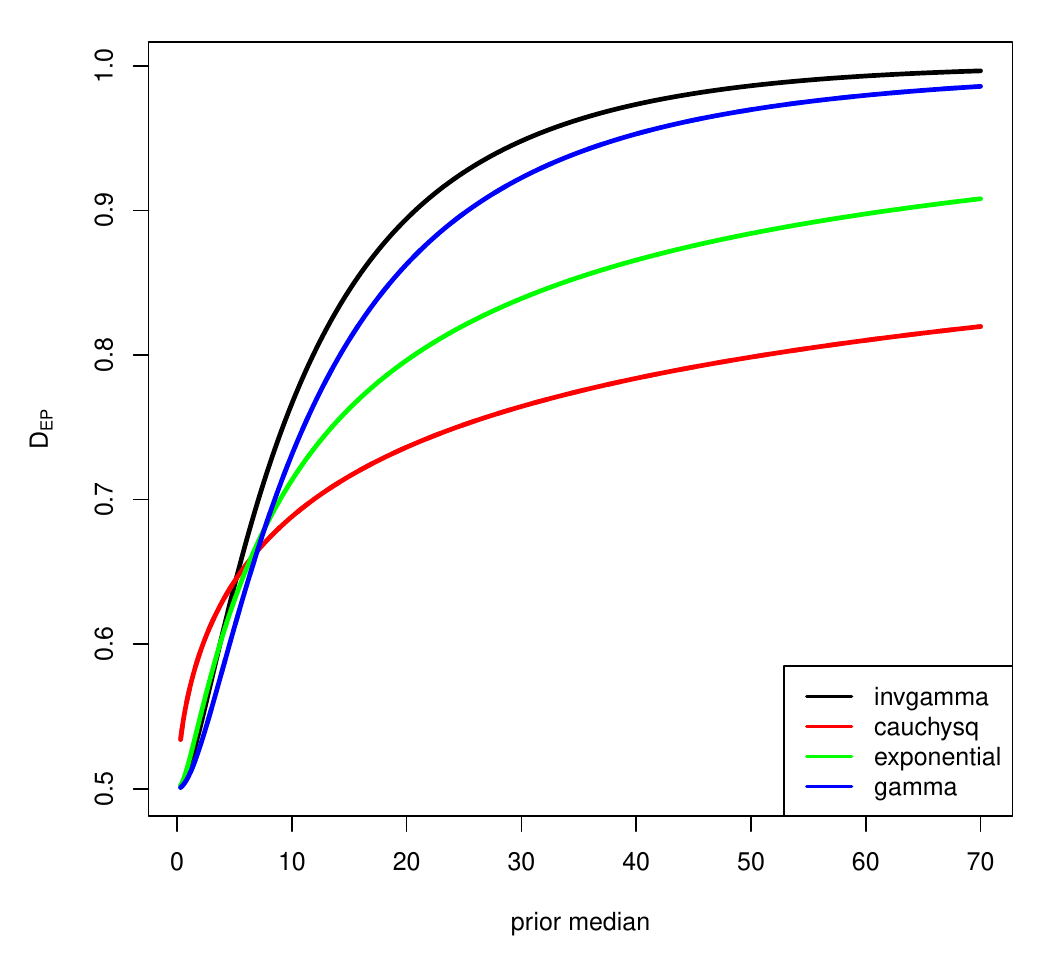}
    \caption{$\dep$ plots for testing the non-centrality parameter $\lambda$ of a $\chi^2_{5}$ random variable. Alternative prior densities on $\lambda$ include an inverse gamma with shape 4.5, a gamma prior with shape 4.5, a squared-Cauchy density, and an exponential prior.} 
    \label{eppX2}
\end{figure}

$\dep$ plots can also be used to compare non-local priors versus null hypotheses.  Fig.~\ref{invMomvMom} displays an $\dep$ plot for normal moment priors \citep{SPL2025}, along with an inverse-moment prior with shape parameter $\nu=9$.
Fig.~\ref{eppX2} displays an $\dep$ plot for testing whether the non-centrality parameter $\lambda$ of a $\chi^2_{5}$ statistic is 0.  Four alternative prior densities are included in the comparison.  These include an exponential prior, a gamma prior with shape parameter 4.5, an inverse gamma prior with shape parameter $\nu/2=4.5$, and a squared-Cauchy prior.  The latter is motivated by a Cauchy prior on a normal mean, which implies a squared-Cauchy prior on a chi-squared non-centrality parameter \citep{Rouder2012}. The density of a squared Cauchy is 
\begin{equation*}
f(x) = \frac{1}{\pi r \sqrt{x} (1+ x/r^2)}.
\end{equation*}

The squared-Cauchy prior provides the strongest support for the true model when concentrated near 0, but comparatively weaker support for more disperse choices of its scale parameter.  Except for small differences near 0, the inverse gamma prior demonstrates higher $\dep$ than the other priors that have the same median.

\section{BFFs based on inverse-moment and inverse gamma densities}

While subjective considerations may motivate alternative choices for the class of prior densities used to construct Bayes Factor Functions (BFFs), our analysis of convergence rates and expected posterior probabilities indicates that inverse-moment and inverse-gamma priors serve as effective default options for classical test statistics. In what follows, we focus specifically on BFFs constructed using inverse-moment priors for noncentrality parameters in $t$ and $z$ tests, and inverse gamma priors for $\chi^2$ and $F$ tests. The implementation of these BFFs requires specifying hyperparameters for each class of prior distributions.

\subsection{Selection of the shape parameter, \texorpdfstring{$\nu$}{nu}}

$\dep$ plots analogous to Figure~\ref{eppT} can also be constructed for $z$-tests, $t$-tests with arbitrary degrees of freedom, and test statistics based on $\chi^2$ and $F$ distributions. In principle, such plots can be combined with subjective prior information to guide the selection of a class of alternative prior distributions to construct BFFs. Alternatively, information from replicated study designs or meta-analyses may inform this selection; this possibility is further explored in Section~\ref{replicated}.

\begin{figure}[ht!]
    \centering
    \includegraphics[width=2.5in]{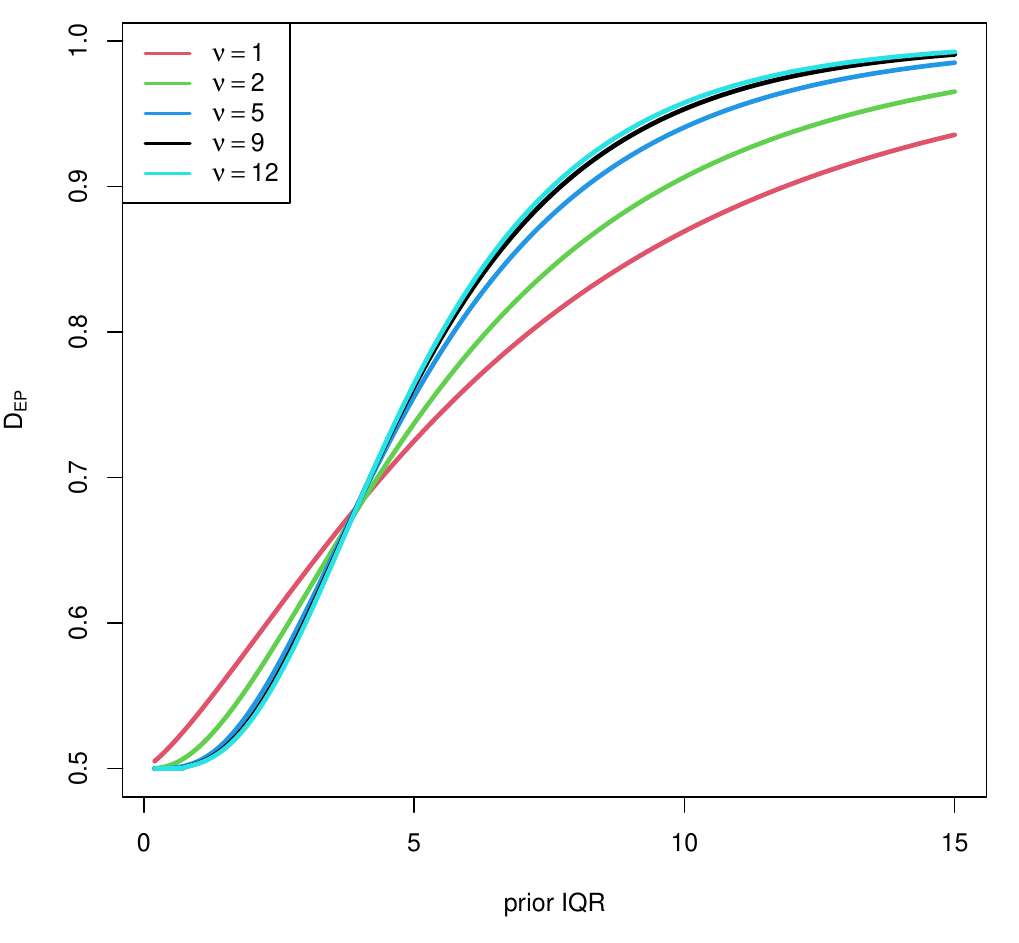}
    \caption{Plots of $\dep$ for various inverse-moment priors as a function of their IQR. } 
    \label{nuComparison}
\end{figure}

Both inverse-moment and inverse-gamma prior families are characterized by shape and scale parameters. Focusing first on inverse-moment priors, $\dep$ plots can inform the choice of the shape parameter $\nu$. Figure~\ref{nuComparison} illustrates such a strategy. As in previous figures, the horizontal axis represents the interquartile range (IQR) of the alternative prior, which is determined by the scale parameter $\tau$.

This plot reveals similar features to those in Figure~\ref{eppT}. For small IQR values, choosing $\nu=1$ yields the highest $\dep$. However, as the IQR increases, the prior mass near the origin decreases more rapidly for higher values of $\nu$, as illustrated in Figure~\ref{imdensities}.

In the absence of subjective prior information or data from replication studies, these findings suggest that setting $\nu=1$ is a conservative default. This choice generally provides higher $\dep$ than larger values of $\nu$ when the alternative prior is concentrated near the null, while also outperforming commonly used local priors as the noncentrality parameter increases.

A potential drawback of selecting $\nu=1$ is that its improved $\dep$ near the null hypothesis comes at the cost of reduced $\dep$ for alternative priors that assign greater mass to larger effect sizes. Moreover, the gains in $\dep$ near the null tend to correspond to $\dep$ values in the range of 0.3–0.7, which may be considered inconclusive from a scientific perspective. 

In many areas of the social and biological sciences, prior knowledge concerning typical standardized effect sizes is available. Investigators often focus on detecting effects within the standardized range 0.2 (small) to 0.8 (large), with 0.5 denoting a medium effect size \citep{Cohen1988}. Consequently, we recommend selecting $\nu$ so that the alternative prior on the noncentrality parameter $\lambda$ places a specified proportion of its mass—say, $\gamma \approx 0.9$—within the range corresponding to standardized effects in $(0.2, 0.8)$, assuming the prior mode for $\omega$ is 0.5.

For $z$ and $t$ tests, this criterion can be implemented as follows. For one sample $z$ and $t$ tests, $\lambda_1 = \sqrt{n} \omega_1$, where $\omega_1$ is a hypothesized standardized effect size. From equation (\ref{invmode}), the value of $\tau$ that places the mode of the prior distribution at $\omega_1=0.5$ is  $\tau_{0.5,\nu} = n (\nu + 1)/8$.  For two-sided tests, we then select $\nu$ such that:
\[
\gamma = \int_{-b}^{-a} i(\lambda \mid n\omega_1^2(\nu+1)/8, \nu) \, d\lambda + \int_a^b i(\lambda \mid n\omega_1^2(\nu+1)/8, \nu) \, d\lambda,
\]
where $a = 0.2\sqrt{n}$ and $b = 0.8\sqrt{n}$ are the $\lambda$ values corresponding to small and large standardized effects.

Using a change of variables, this expression simplifies to:
\[
\gamma = \int_c^d f(y \mid \nu/2, 1) \, dy,
\]
where $c = [2(0.2/0.5)^2]/(\nu+1)$, $d = [2(0.8/0.5)^2]/(\nu+1)$, and $f(\cdot \mid \nu/2, 1)$ is the inverse-gamma density with shape $\nu/2$ and scale 1.

For $\gamma = 0.9$, this condition is satisfied by $\nu = 9$. Because the inverse-moment prior is symmetric around zero, this choice also applies to one-sided $z$ and $t$ tests, as well as two-sample $z$ and $t$ tests when standardized effects are defined as $(\mu_1-\mu_2)/\sigma$.

Applying a similar criterion for $\chi^2$ and $F$ tests yields a similar recommendation, with $\nu = 9$ again emerging as a suitable default.

Accordingly, we recommend $\nu = 9$ as a default choice for the shape parameter for inverse-moment prior densities, and $\nu/2 = 4.5$ for inverse gamma alternative prior densities. Of course, other choices $\nu$ (or other classes of prior densities) can be considered if replicated data or other subjective information suggest otherwise. 

Regardless of how the value of $\nu$ is selected, we treat it as a fixed parameter that governs the dispersion of alternative prior distributions used to construct BFFs. Given a specified value of $\nu$, the parameter $\tau$ is used to center the priors on hypothesized effect sizes $\omega_1$, as illustrated in the following section.

\subsection{Indexing Bayes factor functions by scale parameters}

A central motivation for constructing BFFs is to provide researchers a systematic framework for assessing the strength of evidence supporting a null hypothesis relative to a range of scientifically meaningful alternative hypotheses. Indexing these functions by interpretable quantities is thus of paramount importance.  

The noncentrality parameter of a test statistic provides one potential index, the IQR another. When non-local alternative priors are used, indexing the BFF by a standardized effect size is also possible.  For inverse-moment and inverse gamma priors, this can be accomplished by setting the scale parameter $\tau$ so that the mode of the prior density coincides with a specified standardized effect \citep{Johnson2023}.

For inverse-moment priors with a fixed scale parameter $\nu$, this procedure can be implemented by first expressing a hypothesized value of the non-centrality parameter—denoted $\lambda_1$—as a function of the targeted standardized effect size $\omega_1$, i.e., $\lambda_1 = r(\omega_1)$. We then define $\tau_{\omega_1,\nu}$ implicitly as the value of the prior scale parameter that maximizes the prior density $i(\lambda \con \tau_{\omega_1,\nu})$ at the point $\lambda_1=r(\omega_1)$, that is:
\begin{equation*}
r(\omega_1) = \underset{\lambda}{\arg\max} \ i(\lambda \con \tau_{\omega_1,\nu}). 
\end{equation*} 
Given $\tau_{\omega_1,\nu}$ for a range of $\omega_1$ values, the BFF based on $x$ consists of ordered pairs $(BF(x \con \tau_{\omega_1,\nu}),\omega_1)$.  That is, each point on the BFF provides the Bayes factor obtained by centering the inverse moment prior density on the noncentrality parameter $\lambda_1$ corresponding to $\omega_1$.

For an example of testing a normal mean based on a $t$ statistic, 
the non-centrality parameter of the $t$ statistic is $\lambda = \sqrt{n}\omega$, and the alternative prior density on $\lambda$ is a $i(0,\tau^2,\nu)$ density. This density has maxima at 
$\lambda = \pm \sqrt{2\tau/(\nu+1)}$. For a hypothetical value of $\lambda_1 = \sqrt{n}\omega_1$, this equation can be solved for $\tau_{\omega_1,\nu}$, yielding $\tau_{\omega_1,\nu} = n\omega_1^2(\nu+1)/2$.

Based on this procedure, Table \ref{parset} provides formulae for setting the scale parameter $\tau$ so that the modes of inverse-moment and inverse gamma densities are placed at values of the noncentrality parameter that correspond to a hypothesized standardized effect $\omega_1$. Given the shape parameter $\nu$, inverse-moment densities $i(\lambda \con \tau, \nu)$ are imposed on $z$ and $t$ tests, and inverse gamma densities $ig(\lambda \con \tau, \nu/2)$ are assumed for $\chi^2$ and $F$ tests. This table is similar to Table~1 of \citep{Johnson2023,SPL2025}, which describe $\tau^2$ values for normal-moment and gamma prior densities. 

\begin{table}[ht!]
\Huge
\caption{Default choices of $\tau_{\omega_1,\nu}$}
\label{parset}
\setlength{\tabcolsep}{10pt} 
\renewcommand{\arraystretch}{5} 
\resizebox{\textwidth}{!}{%
\begin{tabular}{lcccccc} \hline \hline
{\bf Test}  & $H_1$ & Prior & Statistic  & \makecell{Standardized \\ Effect ($\omega$)} & \makecell{Non-centrality \\ Parameter ($\lambda$)} & $\tau_{\omega_1,\nu}$ \\ \hline 
\hline

{1-sample z} 
    & $z \sim N(\lambda, 1)$
    & $i(\lambda \mid \tau, \nu)$
    &  {$\frac{\sqrt{n}\bar{x}}{\sigma}$} 
    &  $\frac{\mu}{\sigma}$ 
    & {${\sqrt{n}}{\omega}$}
    & $ \frac{n\omega_1^2(\nu+1)}{2}$ 
           \\  
{1-sample t} 
    & $t \sim T_{k}(\lambda, 1)$
    & $i(\lambda \mid \tau, \nu)$
    & {$\frac{\sqrt{n}\bar{x}}{s}$} 
    &  $\frac{\mu}{\sigma}$ 
    & {${\sqrt{n}}{\omega}$}
    & $ \frac{n\omega_1^2(\nu + 1)}{2}$ 
           \\  
{2-sample z}
    & $z \sim N(\lambda, 1)$
    & $i(\lambda \mid \tau, \nu)$
    & $\frac{\sqrt{n_1 n_2}(\bar{x}_1-\bar{x}_2)}{\sigma\sqrt{n_1+n_2}} $ 
    & $\frac{\mu_1-\mu_2}{\sigma}$ 
    & {$\frac{\sqrt{2n_1 n_2} \omega}{\sqrt{n_1 + n_2}}$}
    & $\frac{n_1 n_2\omega_1^2(\nu + 1)}{n_1+n_2}$ \\ 

{2-sample t} 
    & $t \sim T_{k}(\lambda, 1)$
    & $i(\lambda \mid \tau, \nu)$
    & $\frac{\sqrt{n_1 n_2}(\bar{x}_1-\bar{x}_2)}{s\sqrt{n_1+n_2}} $
    & $\frac{\mu_1-\mu_2}{\sigma}$ 
    & {$\frac{\sqrt{2n_1 n_2} \omega}{\sqrt{n_1 + n_2}}$}
    & $\frac{n_1 n_2\omega_1^2(\nu + 1)}{n_1+n_2}$ \\ 

Multinomial/Poisson 
    & $h \sim \chi^2_k(\lambda)$ 
    & $\textit{IG}\left(\frac{\nu}{2}, \tau\right)$ 
    & $\chi^2_{k} = \sum\limits_{i=1}^k \frac{(n_i-nf_i(\hat{\theta}))^2}{nf_i(\hat{\theta})}$
    & { $ \left( \frac{p_{i}-f_i(\theta)}{\sqrt{f_i(\theta)}} \right)_{k\times 1} $} 
    & {$n \omega' \omega$}
    & {$n \omega_1' \omega_1 \left( \frac{\nu}{2} + 1 \right)$} \\

Linear model
    & $f \sim F_{k,m}(\lambda)$ 
    & $\textit{IG}\left(\frac{\nu}{2}, \tau\right)$
    & $F_{k,n-p} = \frac{(RSS_0-RSS_1)/k}{[(RSS_1)/(n-p)]}$ 
    &{$\frac{\mathbf{C}^{-1}(\mathbf{A}\boldsymbol{\beta}-\mathbf{a})}{\sigma}$} 
    & {$n \omega' \omega$}
    & $n \omega_1'\omega_1\left(\frac{\nu}{2} + 1\right)$
                    \\
              
Likelihood Ratio
    & $h \sim \chi^2_k(\lambda)$ 
    & $\textit{IG}\left(\frac{\nu}{2}, \tau\right)$ 
    & {$ \chi^2_{k} = -2\log\left[\frac{l(\theta_{r0},\hat{\theta_{s}})}{l(\hat{\theta})} \right]$} 
      &{${\bf C}^{-1}(\theta_{r}-\theta_{r0}) $ } 
    & {$n \omega' \omega$}
      & $n\omega_1'\omega_1\left( \frac{\nu}{2} + 1\right)$\\ 
     \hline

\end{tabular}
}
\caption*{For normal mean tests, $\bar{x}$ is the sample mean of $n$ independent $N(\mu,\sigma^2)$ deviates, while $\bar{x}_j$ and $n_j$, $j=1,2$, are the corresponding quantities in two sample tests. The variance is $\sigma^2$. Standard deviations are denoted by $s$ and are the pooled estimate in the two-sample $t$ test. For multinomial/Poisson tests, $f(\boldsymbol{\theta})$ maps an $s \times 1$ vector $\boldsymbol{\theta}$ into a $k \times 1$ probability vector, $k$ the number of cells. The degrees-of-freedom $\delta=k-s-1$; $p_i$ and $n_i$ are cell probabilities and counts, and $n$ is the sum of cell counts.  In the Linear Model, the alternative hypothesis is $\mathbf {A}\boldsymbol{\beta} = \mathbf{ a}$, where $\mathbf{A}$ is a $k\times p$ matrix of rank $k$, $\boldsymbol{\beta}$ is a $p$ vector of regression coefficients, and $\mathbf{a}$ is a $k$ vector. The quantities $RSS_0$ and $RSS_1$ denote the residual sum-of-squares under the null and alternative hypotheses.   For the Likelihood Ratio test, $l(\cdot)$ denotes the likelihood function for $\bt = (\bt_r,\bt_s)$. The $k\times 1$ subvector $\bt_r$ equals $\bt_{r0}$ under the null hypothesis. The MLE of $\bt$ under $H_1$ is $\hat{\bt}$, and the MLE of $\bt_s$ under $H_0$ is $\hat{\bt}_s$.  The matrix ${\bf L}^{-1}$ represents the Cholesky decomposition of the covariance matrix ${\bf C}$ for the tested parameters, scaled to a single observation.  The final column provides the value of $\tau_{\omega_1,\nu}$ that places the mode of the prior density at the non-centrality parameter specified in column 5.}
\end{table}

\section{Examples}\label{sec: sim}

\subsection{Two-sample {\em t} statistics}
The two-sample $t$-test is among the most commonly used test statistics in the social sciences \citep{Cohen1994,MaxwellDelaney2004}.  To illustrate BFFs based upon it, suppose ${x_{ij}}$, $j=1,2$, $i=1,\dots,n_j$ represent conditionally independent draws from $N(\mu_j,\sigma^2)$ distributions, and, following Table~\ref{parset}, we define the standardized effect as $\omega = (\mu_1-\mu_2)/\sigma$, $s^2$ as the pooled estimate of the variance, and $\bar{x}_j$ as the sample mean of each group.  The $t$ statistic is then defined as 
\[ t_{n_1+n_2-2}(\lambda) = \frac{\bar{x}_1-\bar{x}_2}{s\sqrt{1/n_1+1/n_2}},
\]
and has noncentrality parameter 
\[ \lambda = \frac{\sqrt{n_1n_2}\omega}{\sqrt{n_1+n_2}} .\]

Assuming an inverse-moment prior on $\lambda$, the value of $\tau_{\omega_1,\nu}$ that places the prior modes on $\lambda_1 = \sqrt{n_1n_2}\omega_1/\sqrt{n_1+n_2}$  is 
\[ \tau_{\omega_1,\nu} = \frac{n_1 n_2 \omega_1^2(\nu+1)}{n_1+n_2} .\]

For $p \in (0.5,0.025,0.005,0.0025)$ let $q_{1-p,\delta}$ denote the $(1-p)^{\rm th}$ quantiles of standard $t$ statistics on $\delta=n_1+n_2-2$ degrees-of-freedom, i.e., 
\[ P[ T_{\delta}(0) > q_{p,\delta} ] = p. \]

\begin{figure}
\centering
  \includegraphics[width=.75\linewidth]{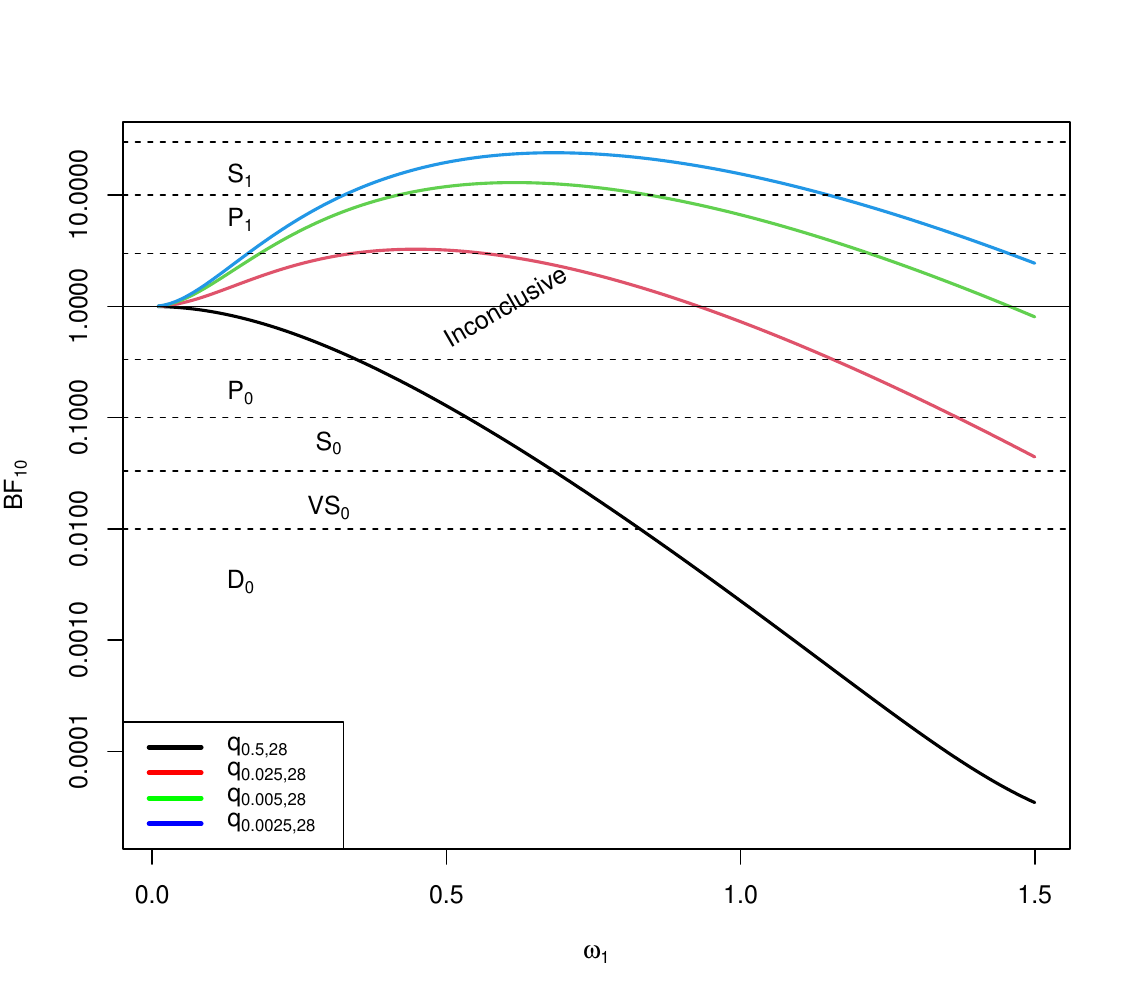}
  \caption{BFFs for $t_{28}$ test statistics defined with inverse-moment priors. The horizontal bands indicate the strength of evidence provided by the Bayes factors for the given value of $\omega_1$. $\rm P_1$ denotes positive evidence ($\rm BF_{10} \in (3-10)$) and $\rm S_1$ strong evidence $(10-30)$. The reciprocal values provide evidence in favor of the null hypothesis. $\rm VS_0$ denotes very strong evidence for the null hypothesis ($\rm BF_{10} \in (1/30-1/100)$), and $\rm D_0$ decisive evidence.}\label{texampleInv}
\end{figure}

Under these assumptions, BFFs for quantiles of a two-sample $t$ statistic with $n_1=n_2 = 15$ and an inverse moment prior with shape parameter $\nu=9$ are displayed Fig.~\ref{texampleInv}. This figure provides a concise summary of evidence in favor of a range of alternative hypotheses based on the observation of quantiles of $t$ statistics drawn from the null distribution. For instance, $t_{15}$ statistics equal to 2.05, 2.76, and 3.05 (two-sided p-values of 0.05, 0.01, 0.005) all provide positive evidence for the alternative hypothesis for certain $\omega_1$ values, with the latter statistics also providing strong evidence for $\omega_1$ values around 0.6. The maximum Bayes factor in favor of the alternative hypothesis was 24.1, achieved for $q_{1-p,28}=3.05 $ with an inverse-moment prior centered on a standardized effect $\omega_1=0.68$, corresponding to a noncentrality parameter of $\lambda_1 = 15\omega_1/\sqrt{30} = 1.86$.  At the other extreme, decisive evidence in favor of the null hypothesis was obtained for $q_{0.5,28} = t_{28} =0$ for all alternative priors centered on
effect sizes greater than $\omega_1=0.83$.  

\begin{figure}[h]
  \includegraphics[width=.75\linewidth]{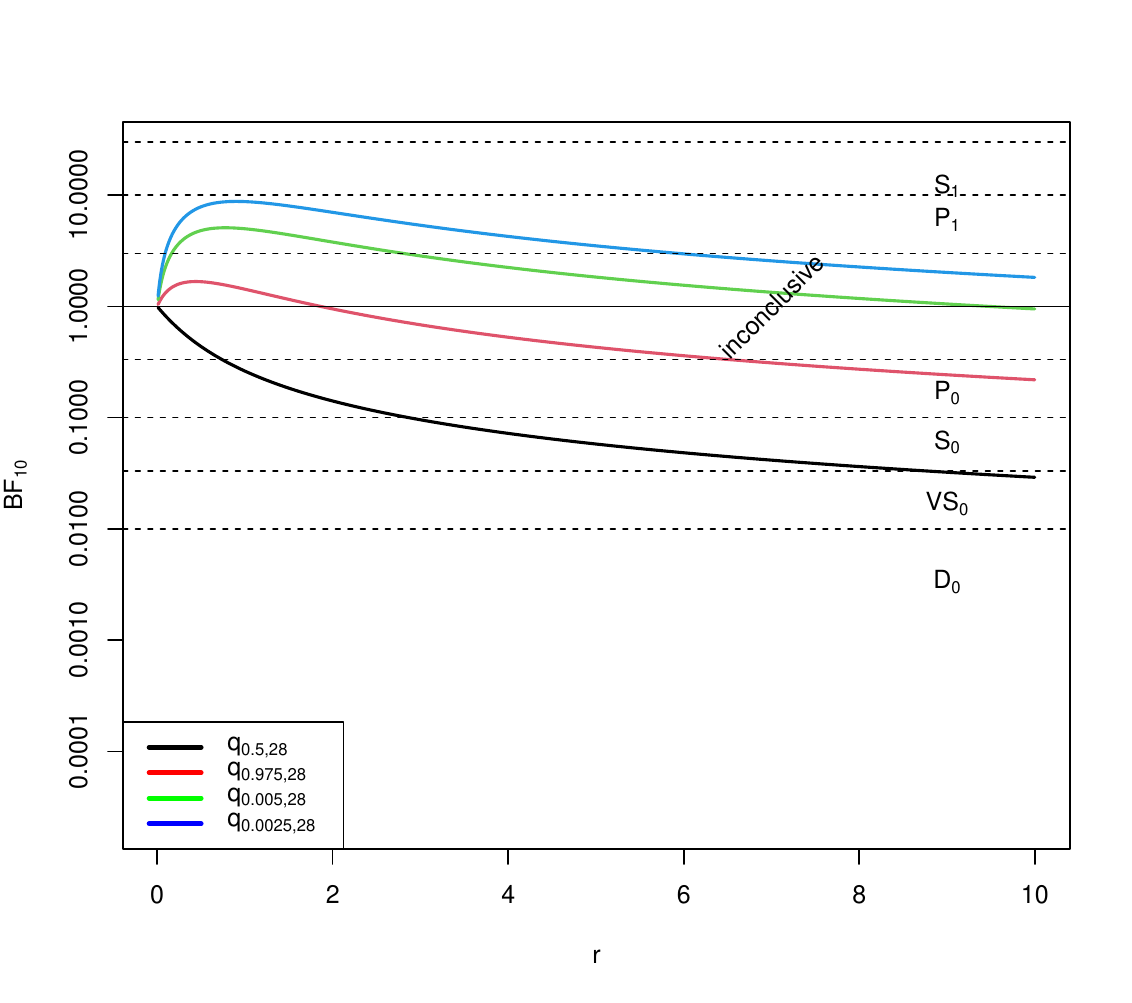}
  \caption{BFFs for $t_{28}$ test statistics defined with JZS priors. The labeling of horizontal bands is similar to Fig.~\ref{texampleInv}, as is the scaling of the vertical axis.}\label{texampleCauchy}
\end{figure}

For comparison, Fig.~\ref{texampleCauchy} provides a similar plot for JSZ priors \citep{Rouder2009}, indexed by the Cauchy scale parameter $r$.  For two-sample $t$ statistics, the noncentrality parameter is $\sqrt{n_1n_2/(n_1+n_2)} \omega$, implying the prior on the $t$ non-centrality parameter is a Cauchy centered on 0 with scale parameter $r\sqrt{n_1n_2/(n_1+n_2)}$.  The BFFs in this plot do not indicate strong evidence in favor of the alternative hypothesis for any values of $r$ for any of the $t_{28}$ quantiles.  Nor do they provide strong evidence in favor of the null hypothesis for commonly recommended values of $r$ (i.e., ($\sqrt{2}/2,2,\sqrt{2}$)).

\subsection{Pearson's \texorpdfstring{$\chi^2$}{chi-squared} statistics}
We next consider Pearson's chi-squared test for goodness-of-fit, as described in Table~\ref{parset}.  Let $\delta=K-s-1$ denote the degrees-of-freedom of the statistic, where $K$ is the number of cells, $s<K-1$ is the dimension of the parameter vector $\theta$ estimated to fit the cell probabilities $p_k$, with $f_k(\hat{\theta})$ the estimate of the cell probability for cell $k$, and $n$ the number of observations. The standardized effect vector $\bfo$ is defined to have components
\[ \bfo =  \left[ \frac{p_k-f_k(\theta)}{f_k(\theta)} \right]_k, \qquad k=1,\dots,K.\]
The non-centrality parameter of the chi-squared statistic for a hypothesized standardized effect $\bfo_1$ is $n\bfo_1'\bfo_1$.  It is not necessary to explicitly compute $\bfo_1$ in the construction of the BFF, only $\bfo'\bfo$ is needed.  Given $\bfo_1$, $\tau_{\omega_1,\nu} =n\bfo_1'\bfo_1(\nu/2+1)$. 
Finally, for $p \in (0.5,0.05,0.01,0.005)$ let $q_{1-p,\delta}$ denote the $(1-p)^{\rm th}$ quantiles of a chi-squared random variable on $\delta$ degrees-of-freedom, i.e., 
\[ P ( X^2_{\delta}(0) > q_{p,\delta} ) = p. \]  For simplicity, we assume $\dim(\theta)=s=0$, corresponding to a simple null hypothesis.

\begin{figure}[h]
  \includegraphics[width=.75\linewidth]{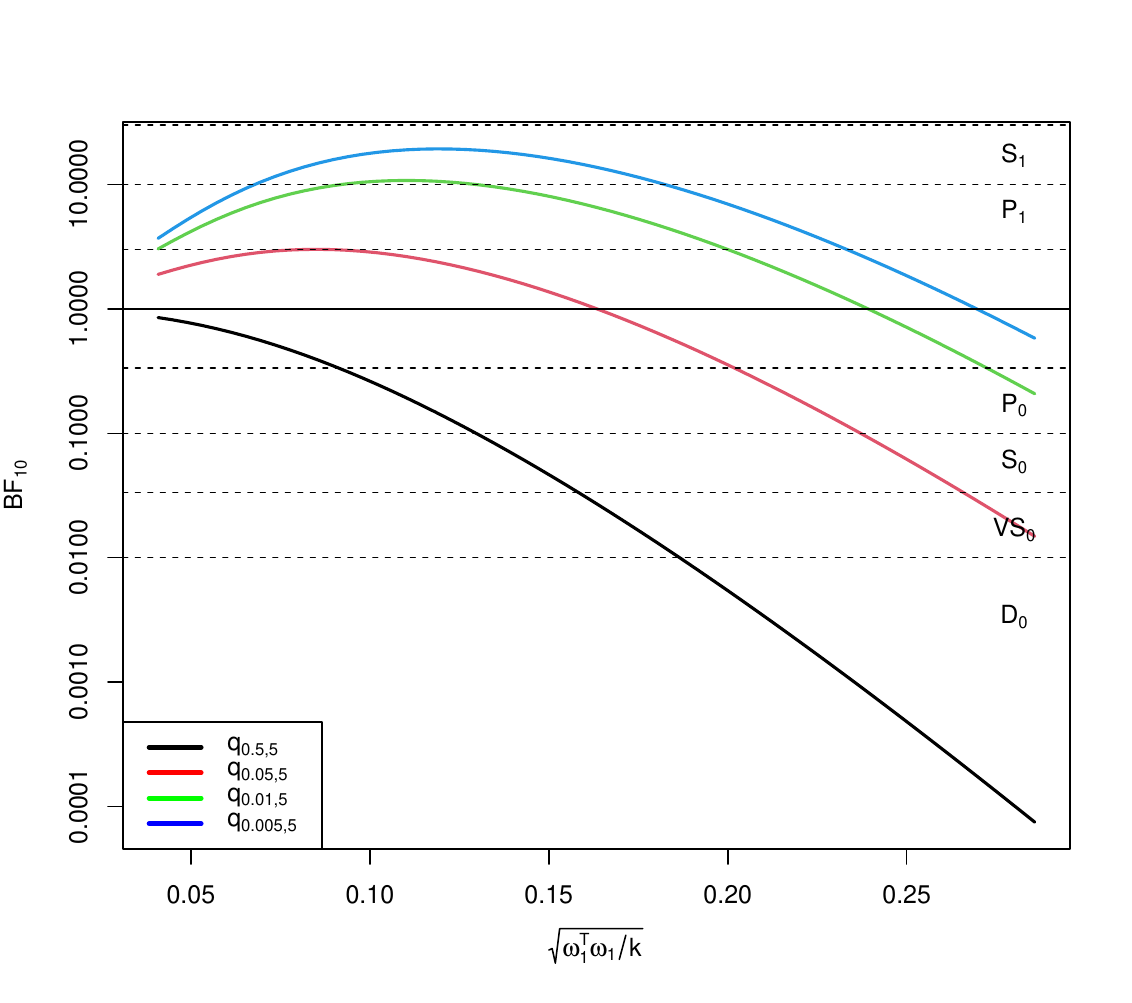}
  \caption{BFFs for $\chi^2_{5}$ test statistics with inverse gamma prior with shape parameter $\nu/2=4.5$. The labeling of horizontal bands is similar to Fig.~8.}\label{x2exInv}
\end{figure}

Fig.~\ref{x2exInv} displays BFFs for chi-squared quantiles on $k=6$ and 5 degrees of freedom. Here, the BFF is indexed by root-mean-square standardized effect size, $\sqrt{\bfo'\bfo/k}$.

In this scenario, when $q_{0.5,5}=4.35$, the Bayes factor yields decisive evidence \newline ($BF_{10}(x)<0.01$) against alternative hypotheses for inverse gamma densities centered on standardized effect vectors for which $\sqrt{\bfo'\bfo/k}> 0.19$. Conversely, when $q_{0.99,5}=15.1$, strong evidence is obtained in favor of the alternative for values of $\sqrt{\bfo'\bfo/k}$ in the interval $ (0.10,0.14)$.

\section{Analysis of replicated experiments and meta-analyses}\label{replicated}
Bayesian methods provide a flexible framework for synthesizing evidence collected from replicated experiments. In such settings, data are often collected under similar designs and test similar hypotheses.  In literature reviews and meta-analyses, evidence provided in support of hypotheses is frequently summarized by a classical test statistic or $p$-value, often without a complete summary of sufficient statistics and other aspects of the experimental design. This fact complicates the aggregation of evidence across studies.

To account for both within- and between-study variability, researchers have increasingly relied on hierarchical Bayesian models. These models introduce latent parameters that represent study-level effects and place priors over their distribution. This framework supports coherent posterior inference that reflects uncertainty in nuisance parameters and provides interpretable estimates for quantities of interest such as overall effects, variance components, and predictive distributions for future studies \citep[e.g.,][]{Higgins2002, Higgins2009, Gelman2013}.

However, in contexts where researchers wish to evaluate specific hypotheses—such as whether an effect is nonzero—Bayesian hypothesis tests using Bayes factors have become an increasingly popular alternative. Bayes factors provide a convenient summary of evidence in support of competing hypotheses. Nonetheless, Bayes factors are not naturally suited to combination across studies in the same manner as likelihood functions because they are not generally multiplicative across independent studies unless further assumptions are imposed about the prior specification and parameter consistency across datasets.

Recent work by \cite{niko2021} has made this limitation explicit, demonstrating that naively multiplying Bayes factors across studies can lead to misleading inferences. If priors are set independently for each study without regard to a common hierarchical structure or shared effect size, then the resulting Bayes factors may reflect disparate prior beliefs and cannot be meaningfully aggregated. On the other hand, if a common effect is assumed across studies, the marginal distribution of the data from all studies must be obtained by integrating the product of likelihood functions with respect to the prior density on the common effect, rather than integrating over the prior for each likelihood independently.  That is, 
\begin{equation*}
\prod_{i=1} \left[ \int f(x_i \con \lambda_i) \pi(\lambda_i\con \nu,\tau,n_i)d\lambda_i \right] \neq \int \prod_i \left[ f(x_i\con \lambda) \right] \,\pi(\lambda\con \nu,\tau)\,d\lambda.
\end{equation*}

In this section, we show how BFFs defined with non-local priors naturally incorporate features of Bayesian hierarchical modeling by centering prior distributions on standardized effects and explicitly accounting for variability in noncentrality parameters and effect sizes across studies. This framework yields Bayes factors that directly support specific hypotheses about standardized effects. While default values for prior dispersion parameters may be employed, we instead propose method-of-moments estimators for these parameters. When BFFs are constructed using inverse-moment or inverse-gamma priors, the resulting estimator of the dispersion parameter $\nu$ is independent of the standardized effect, allowing a single estimate of $\nu$ to be used to define the entire BFF.


\subsection{Empirical Bayes estimates of \(\nu\)}
By imposing inverse-moment and inverse gamma alternative priors on the non-centrality parameters of $z$, $t$, $\chi^2$ and $F$ statistics, it is possible to obtain method-of-moment estimators of the scale parameter $\nu$.  
Heuristically, if $\lambda_i \sim IG( \nu/2,\tau)$, then 
\begin{equation*}
    E(\lambda_i) = \frac{\tau}{\nu/2-1}, \quad Var(\lambda_i) = \frac{\tau^2}{(\nu/2-1)^2(\nu/2-2)} = \frac{E(\lambda_i)^2}{\nu/2-2}
\end{equation*}
for $\nu> 4$. The assumption that $\nu>4$ implies that the variance of the non-centrality parameters is finite. Given a sample of $\{\lambda_i\}$, a simple method-of-moments estimate of $\nu$ is thus 
\begin{equation}\label{mom1}
  4 + \frac{2\bar{\lambda}^2}{s_{\lambda}^2}.   
\end{equation}
Similarly, if $\lambda_i\sim i(\nu,\tau)$, then $\lambda_i^2 \sim IG(\nu/2,\tau)$ and a method-of-moments estimate of $\nu$ can be obtained by substituting the sample mean and variance of $\{\lambda_i^2\}$ into equation~(\ref{mom1}).

In analyzing replicated experiments, the non-centrality parameters of the test statistics are not directly observed. Furthermore, they typically vary across individual experiments $i$ as a function of the sample size, $n_i$.   Thus, it is necessary to standardize the test statistics before calculating method-of-moments estimates.  For example, if $z$ and $t$ tests are based on normal i.i.d.~samples, $\lambda_i = \sqrt{n_i}\omega$, where $\omega$ is the common standardized effect. In that case, dividing the test statistics by $\sqrt{n_i}$ produces statistics that share a common non-centrality parameter. Ignoring terms of order \(O(1/n_i)\), we obtain a closed-form estimator for \(\nu\) given by:
\begin{equation}\label{mom}
    \hat{\nu}_{\text{MOM}} = 4 + \frac{2\bar{x}^2}{S^2}
\end{equation}
where
\[
\bar{x} = \frac{1}{M} \sum_{i=1}^M x_i, \quad S^2 = \frac{1}{M} \sum_{i=1}^M \left( x_i - \bar{x} \right)^2,
\]
and \(x_i\), \(i=1,\dots,M\), denote standardized test statistics. For one-sample $z$ and $t$ tests listed in Table 1, $x_i=z^2_i/n_i$ or $x_i=t_i^2/n_i$, while for the $\chi^2$ and $F$ statistics, $x_i$ is simply the statistic divided by $n_i$. Because $\hat{\nu}_{MOM}>4$, the asymptotic convergence rates cited in Section~3 extend to the resulting aggregated BFFs.   A more formal justification for these estimators is provided in the Supplementary Material.  


\subsection{An example: persistence and conscientiousness data }
An experiment replicated in the Many Labs 3 project (\cite{EBERSOLE201668}) examined the correlation between conscientiousness  
and persistence. Conscientiousness was measured using two items on the Ten Item Personality Inventory (TIPI; \cite{Gosling}) and persistence by the time participants spent solving anagrams, some of which were solvable and others not. The study performed in the Many Labs 3 project represented a modification of a study performed initially in \cite{DeFruyt2000}. Table~\ref{correlations} provides the sample correlations and sample sizes collected in 20 replicated studies.

\begin{table}[ht!]
\centering

\begin{tabular}{|r|r|r|r|r|} \hline
-0.211 (84) & 0.008 (117) & -0.064 (42) & 0.201 (90) & -0.064 (96) \\
0.020 (314) & -0.044 (126) & 0.103 (131) & -0.085 (156) & -0.140 (101)\\ 
 0.024 (118) & 0.004 (139) & 0.142 (179)& 0.060 (117)& -0.020 (240) \\
0.164 (137) &-0.060 (89) & -0.017 (80) & -0.001 (177) & 0.000 (95) \\ 
\hline
\end{tabular}
\caption{Sample correlations (and sample sizes) for the study of the correlation between persistence and conscientiousness.}\label{correlations}
\end{table}
The null hypothesis in this study is that there is no correlation between persistence and conscientiousness measures. To model the sample correlation coefficients, we rescaled Fisher's z-transformation of the sample correlation coefficient \cite{Fisher1915} and defined
\begin{equation*}
    z_i = \frac{\sqrt{n_i-3}}{2} \log\left(\frac{1+\hat{\rho}_i}{1-\hat{\rho}_i} \right).
\end{equation*}
Here, \( \hat{\rho}_i \) denotes the sample correlation coefficient for study \( i \), and \( \rho_i \) denotes the true correlation coefficient. Let \( \boldsymbol{\rho} = (\rho_1, \dots, \rho_{20}) \).

We make the asymptotic assumption that 
\begin{equation*} 
    \qquad z_i {\sim} N\left( \lambda_i, 1 \right), 
\end{equation*}
where 
\[ 
\lambda_i  = \frac{\sqrt{n_i-3}}{2} \log\left(\frac{1+\rho_i}{1-\rho_i} \right).
\]
Let ${\bf z} = \{z_i\}$.

Under the null hypothesis, we assume $H_0:\ \boldsymbol{\rho}=\mathbf{0}$. Under the alternative hypothesis, we define the standardized effect to be
\begin{equation*}
\omega = \frac{1}{2} \log\left(\frac{1+\rho}{1-\rho}\right)
\end{equation*} 
where $\rho$ is the population correlation coefficent,
and assume that $\lambda_i$, given $(\nu,\tau_{\omega,\nu,i})$, are generated independently according to
  \begin{equation*}
    \lambda_i \mid \omega,\nu \sim i(\lambda \mid \tau_{\omega, \nu,i}, \nu), \qquad \mbox{where} \qquad \tau_{\omega, \nu,i} = \frac{(n_i-3)(\nu + 1)\omega}{2}.
\end{equation*} 
This choice of $\tau_{\omega,\nu,i}$ places the modes of the prior distributions on the non-centrality parameters at $\sqrt{n_i-3} \, \omega$.  It is important to note that different prior distributions are assigned to the noncentrality parameters based on associated sample sizes.

For these data, the method-of-moments estimator of $\nu$ was obtained from (\ref{mom}) using the sample mean and variance of the $z_i^2/(n_i-3)$ values and equaled $\hat{\nu}_{\text{MOM}} = 4.75$.

Because the non-centrality parameters $\{\lambda_i\}$ are assumed to be drawn independently from inverse-moment prior densities and the replications of the experiments are also assumed to be conditionally independent, it follows that the Bayes factors based on 20 replications of the experiment, given $\nu$ and $\{\tau_i\}$, can be expressed as the product of Bayes factors from the individual experiments. From the construction of Bayes factors in Section \ref{bffconst}, it follows that the combined Bayes factors are 
\begin{eqnarray*}
    BF_{10}(\mathbf{z} \mid \omega, \nu) &=& 
    \frac{\int \cdots \int \prod_{i=1}^{20} n(z_i \con \lambda_i,1) i(\lambda_i \con \tau_{\omega,\nu,i},\nu) d\lambda_i }{\prod_{i=1}^{20} n(z_i \con 0,1)} \\
    & = &\prod_{i=1}^{20} BF_{10}(z_i|\tau_{\omega,\nu,i},\nu).
\end{eqnarray*}
We stress that this expression is not appropriate if a single non-centrality parameter is assumed to generate all $z_i$ values. In such a case, the joint marginal density of $\bf z$ would not decompose into a product of marginals. Additionally, the assumption of conditional independence among the components of ${\bf z}$ is only reasonable if the prior distribution on ${ \lambda_i }$ are adjusted for sample size so that they are distributed around a common mean. Thus, the mode of the inverse moment prior on $\lambda_i$ should be adjusted based on the values of $n_i$ and $\omega$. 

\begin{figure}[ht!]
    \centering
    \includegraphics[width=0.75\linewidth]{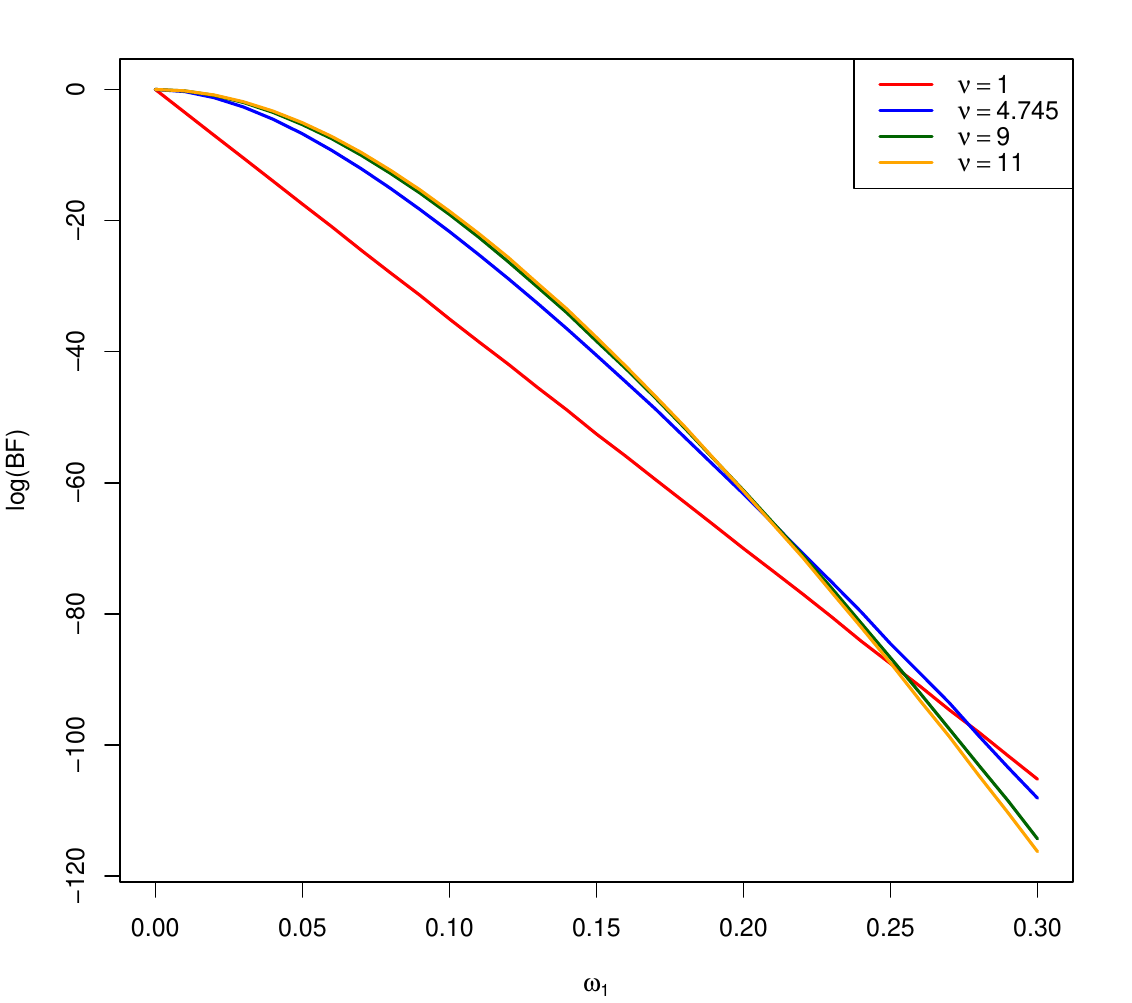}
    \caption{Aggregate BFFs across all sites for testing correlations between Persistence and Conscientiousness for different values of \(\nu\).}
    \label{perscon-aggregate}
\end{figure}

In Figure~\ref{perscon-aggregate}, we display the aggregate BFFs computed for four choices: \(\nu = 1\), \(\nu = \hat{\nu}_{\text{MOM}} = 4.75\), \(\nu = 9\), and \(\nu = 11\). Notably, all four aggregate BFF curves are strictly decreasing in \(\omega_1\), suggesting that the overall evidence in favor of a nonzero correlation diminishes as the effect size increases. All curves demonstrate strong evidence against even small standardized effect sizes. For instance, when \(\hat{\nu}_{\text{MOM}} = 4.75\), the aggregate log Bayes factor at a small effect size of \(\omega_1 = 0.2\) is \(-61.7\).

\begin{figure}[ht!]
    \centering
    \includegraphics[width=0.75\linewidth]{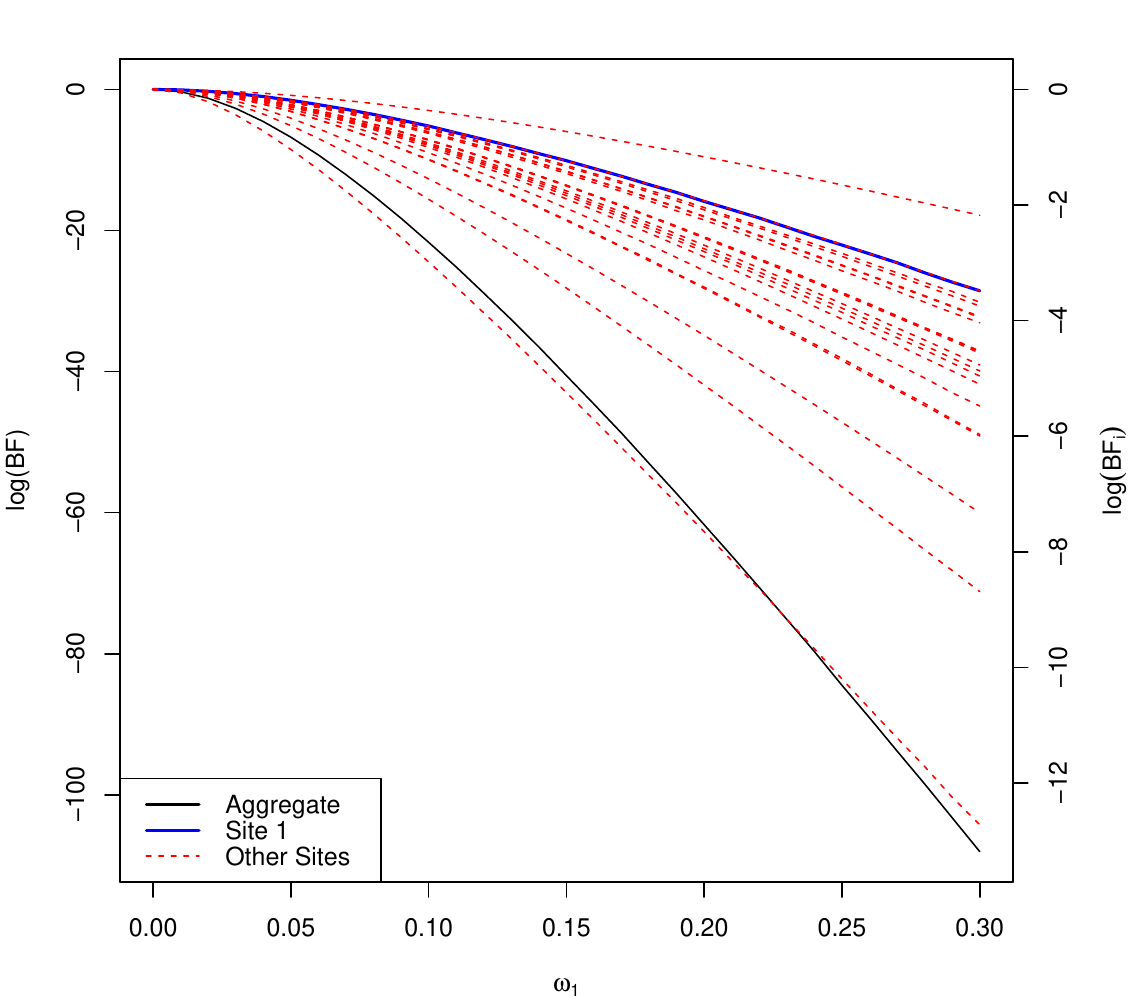}
    \caption{BFFs for testing correlations between Persistence and Conscientiousness using inverse moment priors with \ensuremath{\hat{\nu}_{\text{MOM}} = 4.75}. The scale on the right vertical axis shows the WOE obtained in individual studies, while the scale on the left shows the aggregated weight of evidence.}
    \label{perscon}
\end{figure}

Fig.~\ref{perscon} displays the individual BFFs calculated for the twenty studies reported in Table~\ref{correlations} with \(\nu\) = 4.75.   The blue curve represents the BFF for the first site, with $\hat{\rho}_i=-.211$ and $n=84$. Dashed red curves represent all of the remaining sites, while the solid black curve displays the combined (aggregate) BFF across all sites.  These BFFs were obtained using data from the 20 replications of the persistence-conscientiousness experiment based on a two-sided, one-sample $z$ test.  All individual BFFs are decreasing in \(\omega_1\), and many curves remain far below zero across the range of effect sizes considered, indicating strong support for the null hypothesis. 

As the analysis of $\dep$ in Section 3 suggests, the BFF for $\nu=1$ accumulates evidence more quickly in favor of the null hypothesis in this case than for other values of $\nu$. This occurs because the true standardized effect is very likely to be less than 0.1, although strong evidence for this conclusion is not provided by individual replications of the experiment. Nonetheless, these results suggest that the conservative choice of $\nu=1$ may be preferred for analyzing replicated designs targeted at detecting small standardized effects.

\section{Discussion}

By indexing Bayes factors according to the mode of the prior distribution used to define the alternative hypothesis, BFFs offer informative summaries of the evidence supporting both null and alternative hypotheses across a range of standardized effect sizes.

In this article, the performance of several classes of alternative prior densities used to construct BFFs was compared.  In the case of true null hypotheses, the asymptotic convergence rates of Bayes factors and BFFs for $z$ and $t$ tests using inverse-moment priors were faster than the convergence rates obtained for the same tests with normal moment, JZS or $g$ priors. Similarly, BFFs based on $\chi^2$ and $F$ statistics with inverse gamma priors had better asymptotic convergence rates than those based on gamma, squared Cauchy, and exponential priors on the noncentrality parameters. Expected posterior probability plots generally confirmed the relative performance of these Bayes factors in finite sample settings, except that the JZS and squared Cauchy priors performed similarly (and the squared Cauchy better) for detecting small effect sizes with priors concentrated near the null value. However, the $\dep$ in regions where JZS and squared Cauchy priors perform best occurs when the expected posterior probabilities are less than 0.7 and experimental evidence from a single study is likely to be inconclusive.  Inverse moment priors with $\nu=1$ performed comparably to Cauchy distributions for small effect sizes and provided better performance for priors that were less concentrated around the null.

When testing for the presence of standardized effects in the range of $(0.2,0.8)$ (i.e., small to large standardized effects) based on a single experiment, we recommend a default value of $\nu=9$ for constructing a BFF based on an inverse moment prior for $z$ and $t$ tests, or $\nu/2 = 4.5$ using an inverse gamma prior for $\chi^2$ and $F$ tests.

BFFs constructed with normal moment priors ($z$ and $t$ tests) and gamma priors ($\chi^2$ and $F$ tests) with moderate values of $r$ performed similarly to inverse-moment and inverse gamma priors for comparable values of $\nu$.  Normal moment and gamma priors can also be expressed in closed-form \citep{SPL2025}, making these prior attractive for constructing BFFs to detect moderate to large effect sizes.    

BFFs provide a particularly practical framework for synthesizing evidence in meta-analyses and replicated study designs, especially in scenarios where only classical test statistics or $p$-values are available, or when full likelihood or design information is unavailable. By centering the prior distributions for individual test statistics on noncentrality parameters scaled according to sample size, Bayes factors from separate studies can be coherently aggregated. The resulting cumulative Bayes factor can then be used to construct a BFF that reflects evidence in favor of a range of standardized effects.

The use of method-of-moments estimators for the shape parameters of the inverse-moment and inverse gamma priors further facilitate these analyses. For these prior families, the estimators are independent of the standardized effect, allowing a single estimated dispersion parameter to be used across the full BFF. This approach reduces the subjectivity involved in selecting prior hyperparameters and streamlines the application of BFFs in settings involving replicated experiments and meta-analyses. 

\begin{funding}
The authors acknowledge support from NSF Grant DMS-2311005.
\end{funding}

\bibliographystyle{apalike} 
\bibliography{BASecond}       

\section{Supplementary material: "On Bayes factor functions"}

This supplement provides justification for (a) the convergence rates cited in Sec.~3 of the main article and (b) details on the method-of-moments-type estimators of the shape parameter $\nu$ for inverse-moment and inverse gamma densities based on replicated data.

\section*{Convergence rates}
We use the following lemma to bound the convergence rates of Bayes factors based on test statistics under true null hypotheses.

\begin{lemma}[Integral Inequality for Products of Unimodal Densities]\label{Lemma1}
Let \( f(\lambda) \) and \( g(\lambda) \) be continuous, unimodal probability density functions on \( \mathcal{R} \). Let
\[
\hat{\lambda}_f = \underset{\lambda \in \mathcal{R}}{\arg\max}\, f(\lambda), \quad 
\hat{\lambda}_g = \underset{\lambda \in \mathcal{R}}{\arg\max}\, g(\lambda),
\]
and suppose \( \hat{\lambda}_f \leq c \leq d \leq \hat{\lambda}_g \) for some \( c, d \in \mathcal{R} \). Define
\[
m = \int_{-\infty}^{\infty} f(\lambda) g(\lambda) \, d\lambda.
\]
Then the following inequality holds:
\[
m \leq f(c) + g(d).
\]
\end{lemma}

{\em Proof:}
\begin{eqnarray*}
m &\leq & \int_{-\infty}^{d} f(\lambda) g(\lambda) d\lambda +\int_{c}^{\infty} f(\lambda) g(\lambda)  d\lambda  \\
& \leq & \int_{-\infty}^{d} f(\lambda) g(d) d\lambda +\int_{c}^{\infty} f(c) g(\lambda)  d\lambda\\
&\leq & f(c)+g(d) \qed
\end{eqnarray*}

From the lemma, it follows that if $c=d$ and $f(c)=g(c)$, then $ m\leq 2f(c) = 2 g(c)$.  Also, if $f=g$, then $c=\hf=\hg$ and $m<f(c)$.

As an aside, the following generalization also holds, with similar proof.

\begin{corollary}[Extension to multimodal densities]
Let \( f(\lambda) \) and \( g(\lambda) \) be continuous probability density functions on \( \mathcal{R} \). Suppose for some $c\leq d$ that $f(\lambda)$ is a non-increasing function for $\lambda>c$, and $g(\lambda)$ is a non-decreasing function for $\lambda<d$. Then 
\[ m \leq f(c)+g(d) .\]
\end{corollary}

\begin{lemma}\label{Lemma2}
Let $B(a,b)$ denote the beta function, and let $k,m,n$ denote positive integers. Then
\begin{equation*}
\frac{1}{B\!\left(\tfrac{m}{2}+k,\tfrac{n}{2}\right)}
= \frac{\Gamma\!\left(\tfrac{m}{2}+\tfrac{n}{2}+k\right)}
       {\Gamma\!\left(\tfrac{m}{2}+k\right)\Gamma\!\left(\tfrac{n}{2}\right)}
< \left[1+\frac{n}{2k}\log\!\left(1+\frac{k}{\delta_1-1}\right)\right]^k
  \frac{\Gamma\!\left(\tfrac{m}{2}+\tfrac{n}{2}\right)}
       {\Gamma\!\left(\tfrac{m}{2}\right)\Gamma\!\left(\tfrac{n}{2}\right)},
\end{equation*}
where $\delta_1=\lceil m/2 \rceil$.
\end{lemma}

\begin{proof}
\begin{align*}
\frac{\Gamma\!\left(\tfrac{m}{2}+\tfrac{n}{2}+k\right)}
     {\Gamma\!\left(\tfrac{m}{2}+k\right)\Gamma\!\left(\tfrac{n}{2}\right)}
&= \frac{(\tfrac{m}{2}+\tfrac{n}{2}+k-1)(\tfrac{m}{2}+\tfrac{n}{2}+k-2)\cdots(\tfrac{m}{2}+\tfrac{n}{2})\,\Gamma\!\left(\tfrac{m}{2}+\tfrac{n}{2}\right)}
        {(\tfrac{m}{2}+k-1)(\tfrac{m}{2}+k-2)\cdots(\tfrac{m}{2})\,\Gamma\!\left(\tfrac{m}{2}\right)\Gamma\!\left(\tfrac{n}{2}\right)} \\
&= \Biggl(1+\frac{n}{m+2k-2}\Biggr)\cdots\Biggl(1+\frac{n}{m}\Biggr)
   \frac{\Gamma\!\left(\tfrac{m}{2}+\tfrac{n}{2}\right)}
        {\Gamma\!\left(\tfrac{m}{2}\right)\Gamma\!\left(\tfrac{n}{2}\right)}.
\end{align*}

Furthermore,
\begin{align*}
\frac{1}{k}\sum_{j=1}^k\left(1+\frac{n}{m+2k-2j}\right)
&= 1+\frac{n}{2k}\sum_{j=1}^k \frac{1}{m+2k-2j} \\
&= 1+\frac{n}{2k}\left(\frac{1}{m}+\frac{1}{m+2}+\cdots+\frac{1}{m+2k-2}\right).
\end{align*}

If $m=2\delta_1$ is even, then
\begin{align*}
\frac{1}{m}+\frac{1}{m+2}+\cdots+\frac{1}{m+2k-2}
&= \frac{1}{2}\left(\frac{1}{\delta_1}+\frac{1}{\delta_1+1}+\cdots+\frac{1}{\delta_1+k-1}\right) \\
&\approx \frac{1}{2}\  \left[\log(\delta_1+k-1)-\log(\delta_1-1) \right] \\
&= \frac{1}{2} \ \log\!\left(1+\frac{k}{\delta_1-1}\right) .
\end{align*}

If $m=2\delta_1+1$ is odd, then
\begin{align*}
\frac{1}{m}+\frac{1}{m+2}+\cdots+\frac{1}{m+2k-2}
&< \frac{1}{2}\left(\frac{1}{\delta_1}+\frac{1}{\delta_1+1}+\cdots+\frac{1}{\delta_1+k-1}\right) \\
&\approx \frac{1}{2}\  \left[ \log(\delta_1+k-1)-\log(\delta_1-1) \right] \\
&= \frac{1}{2} \ \log\!\left(1+\frac{k}{\delta_1-1}\right).
\end{align*}

Since the geometric mean of a set of positive real numbers is less than the arithmetic mean, it follows that
\[
\frac{\Gamma\!\left(\tfrac{m}{2}+\tfrac{n}{2}+k\right)}
     {\Gamma\!\left(\tfrac{m}{2}+k\right)\Gamma\!\left(\tfrac{n}{2}\right)}
< \left[1+\frac{n}{2k}\log\!\left(1+\frac{k}{\delta_1-1}\right)\right]^k
  \frac{\Gamma\!\left(\tfrac{m}{2}+\tfrac{n}{2}\right)}
       {\Gamma\!\left(\tfrac{m}{2}\right)\Gamma\!\left(\tfrac{n}{2}\right)}.
\]
\end{proof}

\subsection*{{\em z} tests}

In Theorem 1, let $f(z \mid \lambda, \sigma^2)$ denote the Gaussian density function with mean $\lambda$ and variance $\sigma^2$, and let $g(\lambda \mid \tau,\nu)$ denote an inverse-moment density of the form
\begin{equation}\label{iMom}
  g(\lambda \mid \tau,\nu) = \frac{\tau^{\nu/2}}{\Gamma(\nu/2)} (\lambda^2)^{-\frac{\nu+1}{2}}\exp\left[-\frac{\tau}{\lambda^2}\right].
\end{equation}

The integral of $\int_{-\infty}^\infty fg $ then provides the marginal density of $z$ under the alternative hypothesis that $\lambda \sim g(\lambda \con \tau, \nu)$ (note that $f$ is a density on $\lambda$ given $z$). For the $z$ tests described in Table~1 of the main article, $\tau = an$ for some constant $a> 0$; for convenience of exposition, we set $\tau = n$.

The marginal density of $z$ can be written as the sum of $\int_{-\infty}^0 fg + \int_{0}^\infty fg$. Both integrals are of the same stochastic order, so we bound the integral on the positive real axis, (i.e., $m =\int_0^\infty fg$).

To apply Corollary 1, let $c=d = n^{1/4}$. Then $m$ satisfies
\begin{align*}
    m &\leq  \frac{1}{\sqrt{2\pi}} \exp\left( -\frac{(z - n^{1/4})^2}{2} \right) + \frac{n^{\nu/2}}{\Gamma(\nu/2)} n^{-\frac{\nu+1}{4}}\exp\left(-\frac{n}{n^{1/2}}\right)\\
   & =  O_p[\exp(-an^{1/2})]+ O_p[n^{\nu/4-1/2}\exp(-n^{1/2})]. 
\end{align*}
The value of the density function of $z$ under the null hypothesis (i.e., $f(z\con0,1)$) is $O_p(1)$ under the null, which shows that $BF_{10}(z) = O_p(\exp(-a\sqrt{n}))$ for $0<a<1/2$.

\subsection*{$t$ tests}

Let $t \in \mathbb{R}$, $\lambda \in \mathbb{R}$, and $n \in \mathbb{N}$. The probability density function of the noncentral $t$-distribution with $n$ degrees of freedom and noncentrality parameter $\lambda$ admits the following integral representation:
$$
p(t \mid \lambda) =   \frac{m_0(t)}{2^{\frac{n-1}{2}} \, \Gamma\left( \frac{n+1}{2} \right)} \exp\left( -\frac{n \lambda^2}{2(t^2 + n)} \right) \int_0^\infty y^n \exp\left( -\frac{1}{2} \left( y - \frac{\lambda t}{\sqrt{t^2 + n}} \right)^2 \right) dy,
$$
where $m_0(t)$ denotes the density of the central $t$-distribution with $n$ degrees of freedom.

To apply Corollary 1, define $f(\lambda) = \kappa \frac{p(t\mid \lambda)}{m_0(t)}$, where $\kappa>0$ is a normalizing constant independent of $\lambda$ such that

\begin{align*}
    \kappa \int_\lambda \frac{p(t\mid \lambda)}{m_0(t)} \ d\lambda = 1.
\end{align*}

 We shall prove that $\frac{p(t\mid \lambda)}{m_0(t)}$ is integrable with respect to $\lambda$, i.e., $\kappa$ is a constant bounded away from 0 to show $f(\lambda)$ is a density with respect to $\lambda$. In Corollary~1, we take $c=n^\frac{1}{4}$.

We note that $f(c) = O_p\left(\exp\left(-a\sqrt{n}\right)\right)$, $0<a<\half$, when the conditions $\kappa = O(1)$ and  $ p(t \mid \lambda) = O_p\left(\exp\left(-a\sqrt{n}\right)\right)$ are satisfied, since $m_0(t) = O_p(1) $ under $H_0$. Hence, we verify these two conditions below. 

We now demonstrate that 
$ p(t \mid c) =O_p\left(\exp\left(-a\sqrt{n}\right)\right)$.


To evaluate the integral

$$
I = \int_0^\infty y^n \exp\left( -\frac{1}{2} \left( y - \frac{\lambda t}{\sqrt{t^2 + n}} \right)^2 \right) dy,
$$

we apply the Laplace approximation, $$
\int_0^\infty e^{\psi(y)} dy \approx e^{\psi(y_0)} \sqrt{\frac{2\pi}{-\psi''(y_0)}},
$$
 under the asymptotic regime where $n \to \infty$, $\lambda = n^{1/4}$, and $t = O_p(1)$.

Taking $\psi(y) = n \log y - \frac{1}{2} \left( y - \lambda t / \sqrt{t^2 + n} \right)^2$, each mode $y_0$ satisfies $\psi'(y_0) = 0$, which gives

$$
\frac{n}{y_0} = y_0 - \frac{\lambda t}{\sqrt{t^2 + n}}.
$$

Multiplying both sides by $y_0$ yields the quadratic equation

$$
y_0^2 - \frac{\lambda t}{\sqrt{t^2 + n}} y_0 - n = 0,
$$

whose  roots are

\begin{align}\label{saddle_pt_y}
y_0 = \frac{1}{2} \left( \frac{\lambda t}{\sqrt{t^2 + n}} \pm \sqrt{ \left( \frac{\lambda t}{\sqrt{t^2 + n}} \right)^2 + 4n } \right).
\end{align}

Given that $\lambda = n^{1/4}$ and $\lambda t/\sqrt{t^2+n} = O_p(n^{-1/4})$, we find $y_0 \approx  n^{1/2}$ (the negative root falls outside of the range of integration). Expanding the exponent function at this value gives

$$
\psi(y_0) = n \log y_0 - \frac{1}{2} \left( y_0 - \frac{\lambda t}{\sqrt{t^2 + n}} \right)^2.
$$

It follows that

$$
\psi(y_0) = \frac{n}{2} \log n - \frac{n}{2} + O_p(n^{1/4}).
$$

Moreover, the second derivative satisfies $\psi''(y_0) = -\frac{n}{y_0^2} - 1 = O_p(1)$, since $y_0^2 \approx n$. A Laplace approximation therefore yields \citep{Tierney1989}
\begin{align}
    I = O_p\left(n^{n/2}\exp\left(-n/2\right)\right).
\end{align}

Using Stirling's approximation, we note that, 
\begin{align}
    \Gamma \left(\frac{n+1}{2}\right) &= \left(\frac{n-1}{2}\right)! \nonumber \\ &\approx \sqrt{2\pi} \sqrt{\frac{n-1}{2}} \left(\frac{n-1}{2}\right)^{\frac{n-1}{2}} \exp \left(-\frac{n-1}{2}\right) \nonumber \\
    &= O_p\left(n^{\frac{n}{2}} 2^{-\frac{n}{2}}\exp\left(-\frac{n}{2}\right)\right).
\end{align}

Therefore,
\begin{align}
  p(t \mid \lambda) &=   \frac{m_0(t)}{2^{\frac{n-1}{2}} \, \Gamma\left( \frac{n+1}{2} \right)} \exp\left( -\frac{n \lambda^2}{2(t^2 + n)} \right) \int_0^\infty y^n \exp\left( -\frac{1}{2} \left( y - \frac{\lambda t}{\sqrt{t^2 + n}} \right)^2 \right) dy,\nonumber \\
  &\leq m_0(t) d \frac{\exp\left(-\frac{\sqrt{n}}{2}\right) n^{n/2}\exp(-n/2)}{n^{n/2}\exp(-n/2)2^{-n/2} 2^{n/2}}\nonumber\\
  &= m_0(t) d \exp\left(-\frac{\sqrt{n}}{2}\right),  
\end{align}
where $d>0$ is a constant.
Hence, $\frac{p(t \mid \lambda)}{m_0(t)} = O_p\left(\exp\left(-a\sqrt{n}\right)\right)$ at $c = \lambda = n^{1/4}$ and $0<a<\half$.

We now prove $\kappa = O_p(1)$, bounded away from 0, implying that $f(\lambda)$ can be normalized to be a density in $\lambda$.

Consider the integral,
\begin{align}
&\kappa \int_{-\infty}^\infty f(\lambda) \, d\lambda \nonumber \\&= \frac{\kappa}{2^{\frac{n-1}{2}} \Gamma\left( \frac{n+1}{2} \right)} \int_0^\infty y^n \left( \int_{-\infty}^\infty \exp\left( -\frac{n \lambda^2}{2(t^2 + n)} - \frac{1}{2} \left( y - \frac{\lambda t}{\sqrt{t^2 + n}} \right)^2 \right) d\lambda \right) dy.
\end{align}

The exponent inside the integral is simplified by combining terms:
\begin{align*}
\frac{n \lambda^2}{2(t^2 + n)} + \frac{1}{2} \left( y - \frac{\lambda t}{\sqrt{t^2 + n}} \right)^2 
&= \frac{n \lambda^2}{2(t^2 + n)} + \frac{1}{2} y^2 - \frac{y \lambda t}{\sqrt{t^2 + n}} + \frac{\lambda^2 t^2}{2(t^2 + n)} \\
&= \frac{1}{2} y^2 + \frac{1}{2} \left( \lambda^2 - \frac{2 y \lambda t}{\sqrt{t^2 + n}} \right).
\end{align*}
Completing the square in $\lambda$ yields
\[
\lambda^2 - \frac{2 y \lambda t}{\sqrt{t^2 + n}} = \left( \lambda - \frac{y t}{\sqrt{t^2 + n}} \right)^2 - \frac{y^2 t^2}{t^2 + n}.
\]
Therefore, the exponent can be written
\[
\frac{1}{2} \left( \left( \lambda - \frac{y t}{\sqrt{t^2 + n}} \right)^2 + y^2 \cdot \left(1 - \frac{t^2}{t^2 + n} \right) \right) = \frac{1}{2} \left( \left( \lambda - \frac{y t}{\sqrt{t^2 + n}} \right)^2 + \frac{n y^2}{t^2 + n} \right).
\]

Thus,
\[
\int_{-\infty}^\infty \exp\left( -\frac{n \lambda^2}{2(t^2 + n)} - \frac{1}{2} \left( y - \frac{\lambda t}{\sqrt{t^2 + n}} \right)^2 \right) d\lambda 
= \sqrt{2\pi} \exp\left( -\frac{n y^2}{2(t^2 + n)} \right).
\]

Substituting back, we obtain
\[
\int_{-\infty}^\infty f(\lambda) d\lambda = \frac{\sqrt{2\pi}}{2^{(n-1)/2} \Gamma\left( \frac{n+1}{2} \right)} \int_0^\infty y^n \exp\left( -\frac{n y^2}{2(t^2 + n)} \right) dy.
\]

Letting $c = \frac{n}{2(t^2 + n)}$, we change variables $u = \sqrt{c} y$, so that $dy = \frac{1}{\sqrt{c}} du$, and
\[
\int_0^\infty y^n \exp(-c y^2) dy = \frac{1}{c^{(n+1)/2}} \int_0^\infty u^n e^{-u^2} du = \frac{1}{2 c^{(n+1)/2}} \Gamma\left( \frac{n+1}{2} \right).
\]

Hence,
\[\int_{-\infty}^\infty f(\lambda) d\lambda = \frac{\sqrt{2\pi}}{2^{(n-1)/2} \Gamma\left( \frac{n+1}{2} \right)} \cdot \frac{1}{2 c^{(n+1)/2}} \Gamma\left( \frac{n+1}{2} \right) = \frac{\sqrt{2\pi}}{2^{(n+1)/2} c^{(n+1)/2}}.
\]

This shows that the function
\[
\kappa := \frac{2^{(n+1)/2} c^{(n+1)/2}}{\sqrt{2\pi}} = \frac{2^{(n+1)/2}}{\sqrt{2\pi}} \left( \frac{n}{2(t^2 + n)} \right)^{(n+1)/2}
\]
is the constant that normalizes $f(\lambda)$ into a probability density over $\lambda$.

\bigskip

Assume $t = O(1)$.
Taking logarithms, we have
\begin{align*}
\log \kappa &= \frac{n+1}{2} \log 2 - \frac{1}{2} \log(2\pi) + \frac{n+1}{2} \log\left( \frac{n}{2(t^2 + n)} \right) \\
&= \frac{n+1}{2} \log 2 - \frac{1}{2} \log(2\pi) + \frac{n+1}{2} \left( \log\left( \frac{1}{2} \right) - \frac{t^2}{n} + o(n^{-1}) \right) \\
&= -\frac{1}{2} \log(2\pi) - \frac{n+1}{2} \cdot \frac{t^2}{n} + o(1) = -\frac{1}{2} \log(2\pi) - \frac{t^2}{2} + o(1).
\end{align*}

Therefore,
\[
\kappa = \frac{1}{\sqrt{2\pi}} \exp\left( -\frac{t^2}{2} \right) \cdot (1 + o_p(1)),
\]
and hence \( \kappa = O_p(1) \) as \( n \to \infty \), i.e., the normalization constant remains bounded and bounded away from zero which ensures $f(\lambda)$ is a density in $\lambda$.

Let $g(\lambda)$ denote an inverse moment prior with parameters $(\nu, \tau)$ on $\lambda$. Observing that $f(\lambda)$ is a density in $\lambda$,  using $c=d=n^{1/4}, \tau = \delta n$ and Corollary 1, the Bayes factor against the null hypothesis is,

\begin{align*}
    BF_{10}(t) &= \frac{1}{\kappa}\int_{0}^\infty f(\lambda) g(\lambda) \ d\lambda \\ 
    &\leq \frac{C}{\kappa}\left(\exp\left(-a\sqrt{n}\right) + n^{\frac{\nu}{4}-\frac{1}{2}} \exp\left(-\delta\sqrt{ n}\right)\right) \nonumber \\
    &= O_p(\exp\left(-a\sqrt{n}\right))  
\end{align*}
for some $0<a<\frac{1}{2}, C>0.$

\subsection*{$\chi^2$ tests}

Under the alternative hypothesis $H_1$, the density of the non-central chi-squared distribution with $k$ degrees of freedom and non-centrality parameter $\lambda > 0$ is given by\citep{ananyev2021approximatingmodenoncentralchisquared},
$$
p(h \mid \lambda) = \frac{1}{2} \exp\left( -\frac{h + \lambda}{2} \right) \left( \frac{h}{\lambda} \right)^{\frac{k - 2}{4}} I_{\frac{k - 2}{2}}\left( \sqrt{\lambda h} \right), \quad h > 0,
$$
where $I_\nu(\cdot)$ denotes the modified Bessel function of the first kind of order $\nu$.

To use Lemma 1, we define $f(\lambda) = \kappa \frac{p(h\mid \lambda)}{m_0(h)}$, where $\kappa>0$ is a normalizing constant independent of $\lambda$ such that
\begin{align}
 \int_\lambda \frac{p(h\mid \lambda)}{m_0(h)} \ d\lambda = \frac{1}{\kappa},
\end{align}
and $m_0(h)$ is the density of $h$ under $H_0$. We now prove that $f(\lambda)$ is integrable with respect to $\lambda$ and $\kappa$ is $O_p(1)$. In Lemma 1, we assume $c=n^\frac{1}{2}$.

We note that $f(c) = O_p\left(\exp\left(-\frac{\sqrt{n}}{2}\right)\right)$ if  $\kappa = O(1)$ and $ \ p(h \mid \lambda) =O_p\left(\exp\left(-\frac{\sqrt{n}}{2}\right)\right) $ since $m_0(h) = O_p(1) $ under $H_0$. 

We now verify the two conditions: (i) $ p(h \mid \lambda) =O_p\left(\exp\left(-\frac{\sqrt{n}}{2}\right)\right) $,   (ii) $\kappa = O_p(1)$.

To demonstrate $ {p(h \mid \lambda) =O_p\left(\exp\left(-\frac{\sqrt{n}}{2}\right)\right)} $,we assume $c=\lambda = n^\frac{1}{2}$ and $h = O_p(1)$. From \citep[][p.386, eq.(9)]{Abramowitz1974},

 \begin{align}
   I_{\frac{k - 2}{2}}\left( \sqrt{\lambda h} \right) &\approx \frac{1}{\sqrt{2\pi} } \left(\sqrt{\lambda h}\right)^{-\frac{1}{2}} \exp\left(-\sqrt{\lambda h}\right)\nonumber\\
   & \leq C \exp\left(-n^{1/4}\sqrt{h}\right). 
 \end{align}

This implies, at $c=\lambda = n^\frac{1}{2}$, $p(h \mid \lambda) =O_p\left(\exp\left(-\frac{\sqrt{n}}{2}\right)\right)$. 

Now we prove $\kappa=O_p(1)$ and is bounded away from 0. Consider,
\begin{align}
\int_{-\infty}^{\infty} p\left(h\mid \lambda \right)\, d\lambda 
&= \int_{-\infty}^{\infty} e^{-\lambda/2} e^{-h/2} 
\frac{h^{k/2-1}}{2^{k/2}\Gamma(k/2)} 
\sum_{i=0}^\infty \frac{1}{(k/2)^{(i)}\,i!} 
\left(\frac{\lambda h}{4}\right)^i d\lambda \\
&= e^{-h/2} \frac{h^{k/2-1}}{2^{k/2}\Gamma(k/2)} 
\sum_{i=0}^\infty \frac{\Gamma(i+1)\,2^{i+1}}{(k/2)^{(i)}\,i!} 
\left(\frac{h}{4}\right)^i .
\end{align}
Consider the series
\begin{equation}\label{eq:series}
\sum_{i=0}^\infty \frac{\Gamma(i+1)\,2^{i+1}}{(k/2)^{(i)}\,i!}
\left(\frac{h}{4}\right)^i
= 2\sum_{i=0}^\infty \frac{(h/2)^i}{(k/2)^{(i)}},
\end{equation}
where $(a)^{(i)} = a(a+1)\cdots(a+i-1)$ denotes the rising factorial.

\medskip
\noindent\textbf{Case $k \geq 2$.}  For $k \ge 2$ and each $i>0$, $(k/2)^{(i)} = \prod_{j=0}^{i-1}(k/2+j) \ge \prod_{j=0}^{i-1}(1+j) = i!$. Thus, we obtain
\begin{align*}
2\sum_{i=0}^\infty \frac{(h/2)^i}{(k/2)^{(i)}}
&\leq 2\sum_{i=0}^\infty \frac{(h/2)^i}{i!} \\
&= 2e^{h/2}.
\end{align*}

\medskip
\noindent\textbf{Case $k=1$.}  For $i \ge 1$, $(1/2)^{(i)} = \tfrac{1}{2}\prod_{j=1}^{i-1}(j+\tfrac{1}{2}) \ge \tfrac{1}{2}\prod_{j=1}^{i-1}j = \tfrac{1}{2}(i-1)!$. Therefore,
\begin{align*}
2\sum_{i=0}^\infty \frac{(h/2)^i}{(1/2)^{(i)}}
&= 2 + 2\sum_{i=1}^\infty \frac{(h/2)^i}{(1/2)^{(i)}} \\
&\leq 2 + 4\sum_{i=1}^\infty \frac{(h/2)^i}{(i-1)!} \\
&= 2 + 2h e^{h/2}.
\end{align*}

\medskip
\noindent Combining the two cases, the bound is
\begin{equation}\label{eq:bound}
\sum_{i=0}^\infty \frac{\Gamma(i+1)\,2^{i+1}}{(k/2)^{(i)}\,i!}
\left(\frac{h}{4}\right)^i
\;\leq\;
\begin{cases}
2e^{h/2}, & k \geq 2, \\[6pt]
2 + 2h e^{h/2}, & k = 1.
\end{cases}
\end{equation}
We further note that, $m_0(h) = e^{-h/2} \frac{h^{k/2-1}}{2^{k/2}\Gamma(k/2)}$, i.e., the term multiplied to the sum. Then, it follows that
\begin{align}\label{order_h}
\frac{1}{m_0(h)}\int_{-\infty}^{\infty} p\left(h\mid \lambda \right)\, d\lambda 
&=\sum_{i=0}^\infty \frac{\Gamma(i+1)\,2^{i+1}}{(k/2)^{(i)}\,i!}
\left(\frac{h}{4}\right)^i
\;\leq\;
\begin{cases}
2e^{h/2}, & k \geq 2, \\[6pt]
2 + 2h e^{h/2}, & k = 1.
\end{cases}
\end{align}
Since each summand is nonnegative  and \(h=O_p(1)\), it follows that \\ \(\kappa^{-1}=m_0(h)^{-1}\!\int_{-\infty}^{\infty}p(h\mid\lambda)\,d\lambda\) is bounded away from \(0\) and by \eqref{order_h} we have \(\kappa^{-1}=O_p(1)\). Therefore, for fixed \(k\), \(\kappa=O_p(1)\).

Let $g(\lambda)$ denote an $IG(\frac{\nu}{2},\tau)$ prior on $\lambda$. Then, the Bayes factor against the null is,
\begin{align}
    BF_{10}(h) = \frac{1}{\kappa} \int_{0}^\infty  f(\lambda) g(\lambda) \ d\lambda.
\end{align}

Then, using Lemma 1 and $c = d = \sqrt{n}, \tau = \delta n$, we have,
\begin{align}
    BF_{10}(h) &=\frac{1}{\kappa} \int_{0}^\infty  f(\lambda) g(\lambda) \ d\lambda \nonumber \\
    &\leq \frac{1}{\kappa}\left(f(c) + g(d)\right) \nonumber \\
    &\leq \frac{C}{\kappa}\left(\exp\left(-\frac{\sqrt{n}}{2}\right)+ n^{\frac{\nu}{4}-\frac{1}{2}} \exp\left(-\delta \sqrt{n}\right)\right)\nonumber\\
    &\leq D \exp\left(-a\sqrt{n}\right), \ 0<a<\frac{1}{2}, D,C>0
\end{align}

This implies, $BF_{10}= O_p\left(\exp\left(-a\sqrt{n}\right)\right), \ 0<a<\frac{1}{2}$.

\subsection*{Convergence rates of {\em F} test}

\subsubsection*{Derivation of the pdf of non-central $F$.}
Let \( H \sim \chi^2_m(\lambda) \) and \( Y \sim \chi^2_n \), independent, and define the statistic
\[
F = \frac{H/m}{Y/n} = \frac{n}{m} \cdot \frac{H}{Y}.
\]
We derive the density of \( F \mid \lambda \) in integral form by applying a change of variables and integrating out \( Y \).

The densities of \( H \) and \( Y \) are given respectively by
\[
p_H(h) = \frac{1}{2} \exp\left(-\frac{h + \lambda}{2}\right) \left( \frac{h}{\lambda} \right)^{\frac{m}{4} - \frac{1}{2}} I_{\frac{m}{2} - 1}(\sqrt{\lambda h}) \cdot \mathbf{1}_{\{h > 0\}},
\]
\[
p_Y(y) = \frac{1}{2^{n/2} \Gamma(n/2)} y^{n/2 - 1} \exp(-y/2) \cdot \mathbf{1}_{\{y > 0\}},
\]
where \( I_\nu(\cdot) \) denotes the modified Bessel function of the first kind.

We define a change of variables:
\[
F = \frac{n}{m} \cdot \frac{H}{Y}, \quad Y = Y \quad \Rightarrow \quad H = \frac{m F Y}{n}.
\]
The Jacobian determinant of the transformation is
\[
\left| \frac{\partial(H, Y)}{\partial(F, Y)} \right| = \frac{m Y}{n}.
\]
Thus, the joint density of \( (F, Y) \) is given by
\[
p_{F,Y}(f, y) = p_H\left( \frac{m f y}{n} \right) \cdot p_Y(y) \cdot \frac{m y}{n}.
\]

Substituting the expressions for \( p_H \) and \( p_Y \), we obtain
\begin{align*}
p_{F,Y}(f, y)
&= \frac{1}{2} \exp\left(-\frac{m f y}{2n} - \frac{\lambda}{2} \right)
\left( \frac{m f y / n}{\lambda} \right)^{\frac{m}{4} - \frac{1}{2}}
I_{\frac{m}{2} - 1} \left( \sqrt{ \lambda \cdot \frac{m f y}{n} } \right) \\
&\quad \times \frac{1}{2^{n/2} \Gamma(n/2)} y^{n/2 - 1} \exp(-y/2) \cdot \frac{m y}{n}.
\end{align*}

To obtain the marginal density \( p(f \mid \lambda) \), we integrate out \( y \):
\[
p(f \mid \lambda) = \int_0^\infty p_{F,Y}(f, y) \, dy.
\]

Combining terms and simplifying powers of \( y \), we obtain the integral representation
\begin{align}
p(f \mid \lambda)& =
\frac{m}{2^{n/2 + 1} \, n \, \Gamma(n/2)} \cdot e^{ -\lambda / 2 }
\cdot \left( \frac{m f}{n \lambda} \right)^{\frac{m}{4} - \frac{1}{2}}
\nonumber \\&\times \int_0^\infty y^{n/2 + m/4 - 1/2} \cdot \exp\left( -\frac{y}{2} \left(1 + \frac{m f}{n} \right) \right)
\cdot I_{\frac{m}{2} - 1} \left( \sqrt{ \lambda \cdot \frac{m f y}{n} } \right) dy.
\end{align}

\subsubsection*{Rate of Convergence}

We first show that $p(f \mid \lambda) = O_p(\exp\left(-\frac{\sqrt{n}}{2}\right).$ We analyze the asymptotic behavior of this density in the regime where $n \to \infty$, with fixed $m > 0$, and evaluate the non-centrality parameter $\lambda = n^{1/2}$ in Lemma~1. We also assume that $f = O_p(1)$, i.e., bounded in probability. Our goal is to approximate the integral using Laplace's method.

We focus on the integral term
\begin{align}
I = \int_0^\infty y^{n/2 + m/4 - 1/2} \cdot \exp\left( -\frac{y}{2} \left(1 + \frac{m f}{n} \right) \right)
\cdot I_{m/2 - 1} \left( \sqrt{ \lambda \cdot \frac{m f y}{n} } \right) \, dy.
\end{align}

To approximate the modified Bessel function $I_{m/2 - 1}(z)$ in the integrand, we observe that
\[
z = \sqrt{ \lambda \cdot \frac{m f y}{n} } = \sqrt{ \frac{m f y}{n^{1/2}} },
\]
which implies $z \to 0$ as $n \to \infty$, since $\lambda = c= n^{\frac{1}{2}}$ and $f = O_p(1)$. Therefore, the small-argument expansion of the Bessel function applies:
\[
I_{m/2 - 1}(z) \sim \frac{1}{\Gamma(m/2)} \left( \frac{z}{2} \right)^{m/2 - 1}, \quad z \to 0.
\]

Substituting this approximation, we obtain
\begin{align}
\int_0^\infty y^{n/2 + m/2 - 1}  \exp\left( -\frac{y}{2} \left(1 + \frac{m f}{n} \right) \right)
 \frac{2^{-(m/2 - 1)}}{\Gamma(m/2)}   \cdot \left( \sqrt{\lambda  \frac{m f}{n}} \right)^{(m/2 - 1)}  \, dy.
\end{align}
Let $C := \frac{1}{\Gamma(m/2)}  2^{-(m/2 - 1)}$ be a constant that does not depend on $n$, $\lambda$, or $f$.

Combining the powers of $y$, 
the integral 
\[ I \propto
\left( \sqrt{\lambda  \frac{m f}{n}} \right)^{(m/2 - 1)}  \int_0^\infty y^{n/2 + m/2 - 1}  \exp\left( -\frac{y}{2} \left(1 + \frac{m f}{n} \right) \right) \, dy.
\]

Substituting this back into the full expression for $p(f \mid \lambda)$, we obtain

\begin{align}
p(f \mid \lambda) &\approx \frac{m}{2^{n/2 + 1} \, n \, \Gamma(n/2)} e^{- \lambda / 2}
 \left( \frac{m f}{n \lambda} \right)^{\frac{m}{4} - \frac{1}{2}} 
 \frac{2^{-(m/2 - 1)}}{\Gamma(m/2)}  
 \left( \sqrt{\lambda \cdot \frac{m f}{n} }\right)^{(m/2 - 1)} \nonumber \\
&\quad \times \left( \frac{2}{1 + \frac{m f}{n}} \right)^{n/2 + m/2}  \Gamma\left( \frac{n + m}{2} \right) \nonumber \\
&= e^{-\lambda/2}\,\frac{(m/n)^{m/2}}{B\!\left(\tfrac{m}{2},\tfrac{n}{2}\right)}\,f^{\frac{m}{2}-1}
   \left(1+\frac{mf}{n}\right)^{-\frac{m+n}{2}}.
\end{align}

Using Stirling's approximation for $\Gamma\left(\frac{n}{2}\right)$ and $\Gamma \left(\frac{n+m}{2}\right)$, and substituting $\lambda = n^{1/2}$, we obtain
\begin{equation}\label{eqn: order_c1}
  p(f \mid \lambda) = O\left(  \exp\left( -\frac{\sqrt{n}}{2} \right) \right).  
\end{equation}

Now, similar to the previous proofs, we define $f(\lambda) = \frac{1}{\kappa} \frac{p(f \mid \lambda)}{m_0(f)}$. We prove $\int_0^\infty \frac{p(f \mid \lambda)}{m_0(f)} \ d\lambda= O_p(1)$ and bounded away from 0.

The density of the non-central $F$ distribution with numerator degrees of freedom $m$, denominator degrees of freedom $n$, and non-centrality parameter $\lambda$ is given by
\begin{align}
p(f \mid \lambda) &=
\left(\frac{m}{n}\right)^{\frac{m}{2}} \left(\frac{n}{n+mf}\right)^{\frac{n+m}{2}} f^{\frac{m}{2}-1} \nonumber \\
&\times \sum_{k=0}^\infty \frac{\exp\left(-\frac{\lambda}{2}\right)\left(\frac{\lambda}{2}\right)^k \Gamma\left(\frac{m+n}{2}+k\right)}{k! \Gamma\left(\frac{n}{2}\right)\Gamma\left(\frac{m}{2}+k\right)} \left(\frac{m}{n}\right)^k \left(\frac{nf}{n+mf}\right)^k \nonumber \\
& = \left(\frac{m}{n}\right)^{\frac{m}{2}} \left(\frac{n}{n+mf}\right)^{\frac{n+m}{2}} f^{\frac{m}{2}-1} \frac{1}{\beta\left(\frac{m}{2}, \frac{n}{2}\right)}\nonumber \\
& \times \sum_{k=0}^\infty \frac{\exp\left(-\frac{\lambda}{2}\right)\left(\frac{\lambda}{2}\right)^k \left(\frac{m+n}{2}\right)^{(k)}}{k! \Gamma\left(\frac{n}{2}\right)\left(\frac{m}{2}\right)^{(k)}} \left(\frac{m}{n}\right)^k \left(\frac{nf}{n+mf}\right)^k. 
\label{eq:noncentralF}
\end{align}

The density of the central $F$ distribution is obtained by setting $\lambda = 0$ in \eqref{eq:noncentralF}, yielding
\begin{equation*}
m_0(f) =
 \left(\frac{m}{n}\right)^{\frac{m}{2}} \left(\frac{n}{n+mf}\right)^{\frac{n+m}{2}} f^{\frac{m}{2}-1} \frac{1}{\beta\left(\frac{m}{2}, \frac{n}{2}\right)}.
\label{eq:centralF}
\end{equation*}

The ratio of the non-central to central $F$ densities simplifies to
\begin{equation*}
\frac{p(f \mid \lambda)}{m_0(f)} =
\sum_{k=0}^\infty \frac{\exp\left(-\frac{\lambda}{2}\right)\left(\frac{\lambda}{2}\right)^k \left(\frac{m+n}{2}\right)^{(k)}}{k! \Gamma\left(\frac{n}{2}\right)\left(\frac{m}{2}\right)^{(k)}} \left(\frac{m}{n}\right)^k \left(\frac{nf}{n+mf}\right)^k. 
\label{eq:Fratio}
\end{equation*}

Define $\kappa$ such that $\int \frac{p(f \mid \lambda)}{m_0(f)} \ d\lambda = \frac{1}{\kappa}. $  Let $a = \int_{-\infty}^{\infty} p\left(f \mid \lambda\right) \ d\lambda$.  Then,

\begin{align*}
a&=\int_{-\infty}^{\infty} p\left(f \mid \lambda \right)\, d\lambda \nonumber \\
&= \int_{-\infty}^{\infty} \sum_{k=0}^\infty
 \frac{ e^{-\lambda/2} (\lambda/2)^k}{B\!\left(\tfrac{n}{2},\tfrac{m}{2}+k\right)k!} 
 \left(\frac{m}{n}\right)^{\tfrac{m}{2}+k}
 \left(\frac{n}{n+mf}\right)^{\tfrac{m}{2}+\tfrac{n}{2}+k} 
 f^{\tfrac{m}{2}-1+k} \, d\lambda \\
&=  c(m,n,f) \int_{-\infty}^{\infty}  \sum_{k=0}^\infty
 \frac{ e^{-\lambda/2} (\lambda/2)^k}{B\!\left(\tfrac{n}{2},\tfrac{m}{2}+k\right)k!} 
 \left(\frac{m}{n}\right)^{k}\left(\frac{n}{n+mf}\right)^{k} f^{k} \, d\lambda ,
\end{align*} 
where
\begin{equation*}
 c(m,n,f) = \left(\frac{m}{n}\right)^{\tfrac{m}{2}}
 \left(\frac{n}{n+mf}\right)^{\tfrac{m}{2}+\tfrac{n}{2}} 
 f^{\tfrac{m}{2}-1}.
\end{equation*}

Integrating out $\lambda$ yields
\begin{align}\label{eqn:marg_f}
\int_{-\infty}^{\infty} p\left(f \mid \lambda \right)\, d\lambda 
&= c(m,n,f)  \sum_{k=0}^\infty
 \frac{1}{B\!\left(\tfrac{n}{2},\tfrac{m}{2}+k\right)} 
 \left(\frac{m}{n}\right)^{k}
 \left(\frac{n}{n+mf}\right)^{k} f^{k} 
 \end{align}
 \\
By Lemma \ref{Lemma2},
 \begin{align}
\int_{-\infty}^{\infty} p\left(f \mid \lambda \right)\, d\lambda&< c(m,n,f) \frac{\Gamma\!\left(\tfrac{m}{2}+\tfrac{n}{2}\right)}{ \Gamma\!\left(\tfrac{m}{2}\right)\Gamma\!\left(\tfrac{n}{2}\right)}\nonumber \\
&\quad \times \sum_{k=0}^\infty
 \left\{ \left[ 1+\frac{n}{2k} \log\!\left(1+\frac{k}{\delta_1-1}\right) \right] 
 \frac{m f}{n+mf}\right\}^k,  \label{prodsum}
\end{align}
where $\delta_1 = \lceil m/2 \rceil$.
We note that, $m_0(f) = c(m,n,f) \frac{\Gamma\!\left(\tfrac{m}{2}+\tfrac{n}{2}\right)}{ \Gamma\!\left(\tfrac{m}{2}\right)\Gamma\!\left(\tfrac{n}{2}\right)}$. Hence, equation \eqref{prodsum} implies 
\begin{align}\label{order_f}
   \frac{1}{m_0(f)} \int_{-\infty}^{\infty} p\left(f \mid \lambda \right)\, d\lambda &< \sum_{k=0}^\infty
 \left\{ \left[ 1+\frac{n}{2k} \log\!\left(1+\frac{k}{\delta_1-1}\right) \right] 
 \frac{m f}{n+mf}\right\}^k.
\end{align}

For large $n$, the sum in equation \eqref{order_f} represents a power series in $k$ with terms of order $[ O_p(1/n)+O_p(1/k)]^k$.   Under the null, $m_0(f) = O_p(1)$, and from equation \eqref{eqn:marg_f}, the integral is bounded away from 0.  Hence, 
 $\int_0^\infty \frac{p(f \mid \lambda)}{m_0(f)} \ d\lambda = \frac{1}{\kappa} = O_p(1)$ and $\kappa = O_p(1)$.

 This proves that $f(\lambda)$ can be normalized to be a density and that the normalization constant is $O_p(1)$.
 
Similar to the $\chi^2$ test, define $g(\lambda)$ to be an $IG\left(\frac{\nu}{2}, \tau\right)$ prior on $\lambda$. Using  $c = d = \sqrt{n}, \tau = \delta n$ and Lemma \ref{Lemma1},

\begin{align}
    BF_{10}(f) &= \frac{1}{\kappa}\int_\lambda f(\lambda) g(\lambda) \ d\lambda \nonumber \\
    &\leq \frac{1}{\kappa}\left(f(c) + g(d)\right) \nonumber\\
    & =\frac{C}{\kappa}\left(\exp\left(-\frac{\sqrt{n}}{2}\right) + n^{\frac{\nu}{2} - \frac{1}{4}} \exp\left(-\delta \sqrt{n}\right)\right) \nonumber \\
    &= O_p\left(\exp\left(-a\sqrt{n}\right)\right), \ 0<a<\frac{1}{2}, \ C>0.
\end{align}

\section*{Empirical Bayes Estimate of \texorpdfstring{$\nu$}{nu}}
\subsection*{Choice of $\nu$ for a $z$-test}
Under $H_1,$ $Z_i \mid \lambda_i \sim \mathcal{N}(\lambda_i, 1)$ and $\lambda_i \sim i(\lambda \mid \tau, \nu)$ for $i = 1,2,..., M$.  
Define \(X_i = Z_i^2\), and \(\lambda_i^2 = \theta_i\)\\  
Hence, \(X_i \mid \theta_i \sim \chi_1^2(\theta_i)\) and $\theta_i \sim IG\left(\frac{\nu}{2}, \tau_i\right)$

\begin{equation*}
\mathbb{E}[X_i] = \mathbb{E}_{\theta_i}[\mathbb{E}[X_i \mid \theta_i]] = \mathbb{E}[1 + \theta_i] = 1 + \mathbb{E}[\theta_i] = 1 + \frac{2\tau_i}{\nu - 2} = 1 + \frac{n_i\omega^2(\nu +1)}{\nu - 2}
\end{equation*}

This implies,

\begin{equation*}
\mathbb{E}\left[\frac{X_i}{n_i}\right] = \frac{1}{n_i} + \frac{\omega^2(\nu +1)}{\nu - 2}
\end{equation*}

Now,

\begin{align*}
\mathrm{Var}(X_i) &= \mathrm{Var}_{\theta_i}(\mathbb{E}[X_i \mid \theta_i]) + \mathbb{E}_{\theta_i}[\mathrm{Var}(X_i \mid \theta_i)] \nonumber \\
&= \mathrm{Var}(\theta_i)  + 4\mathbb{E}[\theta_i] \nonumber + 2 \\
&= \frac{\tau_i^2}{\left( \frac{\nu}{2} - 1 \right)^2 \left( \frac{\nu}{2} - 2 \right)}  + \frac{8 \tau_i}{\nu - 2} \nonumber  + 2\\
&= \frac{2n_i^2\omega^4(\nu + 1)^2}{\left( \nu-2 \right)^2 \left(\nu -4 \right)} + \frac{4 n_i\omega^2(\nu +1)}{\nu - 2} + 2
\end{align*}

Hence,
\begin{equation*}
\mathrm{Var}\left(\frac{X_i}{n_i}\right) = \frac{2\omega^4(\nu + 1)^2}{\left( \nu-2 \right)^2 \left(\nu -4 \right)}  + \frac{4\omega^2(\nu +1)}{n_i(\nu - 2)} + \frac{2}{n_i^2}
\end{equation*}

Ignoring the \( O\left(\frac{1}{n_i}\right) \) terms, the MOM estimator of $\nu$ is,

\begin{equation*}
\hat{\nu}_{\text{MOM}} = 4 + \frac{2\bar{x}^2}{S^2}
\end{equation*}

where $z_i'$s are the observed test statistics, $x_i = z_i^2$, $i = 1,2,\ldots,M$, \[\Bar{x} = \frac{1}{M} \sum_{i=1}^M \left(\frac{x_i}{n_i}\right)\] and \[S^2 = \frac{1}{M}\sum_{i=1}^M \left(\frac{x_i}{n_i} - \frac{1}{M} \sum_{i=1}^M \left(\frac{x_i}{n_i}\right)\right)^2.\]

\subsection*{Choice of $\nu$ for a $t$-test}
Under $H_1,$ $t_i \mid \lambda_i \sim T_{\mu_i}(\lambda_i)$ and $\lambda_i \sim i(\lambda \mid \tau, \nu)$ for $i = 1,2,..., M$.  
Define \(X_i = t_i^2\), and \(\lambda_i^2 = \theta_i\)\\  
Hence, $X_i \mid \theta_i \sim F_{1, \mu_i}(\theta_i)$ and $\theta_i \sim IG\left(\frac{\nu}{2}, \tau_i\right)$

\begin{equation*}
\mathbb{E}[X_i] = \mathbb{E}_{\theta_i}[\mathbb{E}[X_i \mid \theta_i]] = \frac{\mu_i}{\mu_i - 2}(1 + \mathbb{E}[\theta_i]) = \frac{\mu_i}{\mu_i - 2} \left(1 + \frac{2 \tau_i}{\nu - 2} \right).
\end{equation*}

This implies,
\begin{equation*}
\mathbb{E}\left[\frac{X_i}{n_i}\right] = \frac{\mu_i}{\mu_i - 2} \left( \frac{1}{n_i} + \frac{ \omega^2 (\nu + 1)}{\nu - 2} \right).
\end{equation*}

Now,
\begin{equation*}
\mathrm{Var}(X_i) = \mathbb{E}_{\theta_i}[\mathrm{Var}(X_i \mid \theta_i)] + \mathrm{Var}_{\theta_i}(\mathbb{E}[X_i \mid \theta_i]).
\end{equation*}

\begin{equation*}
\mathrm{Var}(X_i) = \left( \frac{\mu_i}{\mu_i - 2} \right)^2 \mathrm{Var}(\theta_i)
+ \frac{2\mu_i^2}{(\mu_i - 4)(\mu_i - 2)^2} \left( \mathbb{E}[\theta_i^2] + 2\mathbb{E}[\theta_i] + 1  + (\mu_i -2)(1+2\mathbb{E}[\theta_i])\right).
\end{equation*}

This implies,
\begin{align}
\mathrm{Var}\left(\frac{X_i}{n_i}\right)
&= \left( \frac{\mu_i}{\mu_i - 2} \right)^2 
\cdot \frac{2 \omega^4 (\nu + 1)^2}{(\nu - 2)^2(\nu - 4)} \nonumber \\[8pt]
&\quad + \frac{2\mu_i^2}{(\mu_i - 4)(\mu_i - 2)^2} \cdot \Bigg[
\frac{2 \omega^4 (\nu + 1)^2}{(\nu - 2)^2(\nu - 4)} 
+ \frac{\omega^4 (\nu + 1)^2}{(\nu - 2)^2} \nonumber \\[8pt]
&\qquad + \frac{4 \omega^2 (\nu + 1)}{n_i (\nu - 2)} 
+ \frac{1}{n_i^2}
+ (\mu_i - 2)\left( \frac{8 \omega^2 (\nu + 1)}{2n_i(\nu - 2)} + \frac{1}{n_i^2} \right)
\Bigg]
\end{align}

Hence, assuming \( \mu_i = O (n_i) \)
and ignoring the \( O\left(\frac{1}{n_i}\right) \) terms, the MOM estimator of $\nu$ is,

\begin{equation*}
\hat{\nu}_{\text{MOM}} = 4 + \frac{2\bar{x}^2}{S^2}
\end{equation*}

where $t_i'$s are the observed test statistics, $x_i = t_i^2$, $i = 1,2,\ldots,M$, \[\Bar{x} = \frac{1}{M} \sum_{i=1}^M \left(\frac{x_i}{n_i}\right)\] and \[S^2 = \frac{1}{M}\sum_{i=1}^M \left(\frac{x_i}{n_i} - \frac{1}{M} \sum_{i=1}^M \left(\frac{x_i}{n_i}\right)\right)^2.\]

\subsection*{Choice of $\nu$ for a $\chi^2$-test}
Under \( H_1 \), \( H_i \mid \lambda_i \sim \chi^2_k(\lambda_i) \), \( \lambda_i \sim \text{IG}\left( \frac{\nu}{2}, \tau \right) \)

\begin{equation*}
\mathbb{E}[H_i] = \mathbb{E}_{\lambda_i} \left[ \mathbb{E}[H_i \mid \lambda_i] \right] 
= k + \frac{n_i \tilde{\omega}^2 (\nu + 2)}{\nu - 2}
\end{equation*}
Hence,
\begin{equation*}
\mathbb{E}\left[\frac{H_i}{n_i}\right] = \frac{k}{n_i} + \frac{\tilde{\omega}^2 (\nu + 2)}{\nu - 2}
\end{equation*}\\

Now,
\begin{align*}
\mathrm{Var}(H_i) 
&= \mathrm{Var}_{\lambda_i}\left( \mathbb{E}[H_i \mid \lambda_i] \right) + \mathbb{E}_{\lambda_i}\left( \mathrm{Var}(H_i \mid \lambda_i) \right) \\[6pt]
&= \frac{2 n_i^2 \tilde{\omega}^4 (\nu + 2)^2}{(\nu - 2)^2 (\nu - 4)} + 2k + \frac{4 n_i \tilde{\omega}^2 (\nu + 2)}{\nu - 2}
\end{align*}
Hence, 
\begin{equation*}
\mathrm{Var}\left( \frac{H_i}{n_i} \right) 
= \frac{2 \tilde{\omega}^4 (\nu + 2)^2}{(\nu - 2)^2 (\nu - 4)} 
+ \frac{2k}{n_i^2} 
+ \frac{4 \tilde{\omega}^2 (\nu + 2)}{n_i (\nu - 2)}
\end{equation*}

Ignoring the \( O\left(\frac{1}{n_i}\right) \) terms, the MOM estimator of $\nu$ is,

\begin{equation*}
\hat{\nu}_{\text{MOM}} = 4 + \frac{2 \bar{x}^2}{S^2}
\end{equation*}

where $H_i$s are the observed test statistics, $i = 1,2,\ldots,M$, \[\Bar{x} = \frac{1}{M} \sum_{i=1}^M \left(\frac{H_i}{n_i}\right)\] and \[S^2 = \frac{1}{M}\sum_{i=1}^M \left(\frac{H_i}{n_i} - \frac{1}{M} \sum_{i=1}^M \left(\frac{H_i}{n_i}\right)\right)^2.\]

\subsection*{Choice of $\nu$ for an $F$-test}
Under \( H_1 \), \( F_i \mid \lambda_i \sim F_{k, m_i}(\lambda_i), \quad \lambda_i \sim \text{IG}\left( \frac{\nu}{2}, \tau \right) \)

\begin{equation*}
\mathbb{E}[F_i] 
= \mathbb{E}\left[ \mathbb{E}[F_i \mid \lambda_i] \right] 
= \mathbb{E}\left[ \frac{m_i (k + \lambda_i)}{k(m_i - 2)} \right] 
= \frac{m_i}{m_i - 2} + \frac{n_i m_i \tilde{\omega}^2 (\nu + 2)}{k (m_i - 2)(\nu - 2)}
\end{equation*}
Hence, 
\begin{equation*}
\mathbb{E}\left[\frac{F_i}{n_i}\right] = \frac{m_i}{n_i(m_i - 2)} 
+ \frac{m_i \tilde{\omega}^2 (\nu + 2)}{k (m_i - 2)(\nu - 2)}
\end{equation*}\\
Now, 
\begin{align*}
\mathrm{Var}(F_i) 
&= \mathrm{Var}_{\lambda_i}\left( \mathbb{E}[F_i \mid \lambda_i] \right) + \mathbb{E}_{\lambda_i}\left( \mathrm{Var}(F_i \mid \lambda_i) \right) \\[6pt]
&= \left( \frac{m_i}{m_i - 2} \right)^2 \cdot \mathrm{Var}(\lambda_i)
+ \frac{2 m_i^2}{k^2 (m_i - 2)^2 (m_i - 4)} 
\left[ 
(k + \mathbb{E}[\lambda_i])^2 + (m_i - 2)(k + 2\mathbb{E}[\lambda_i])
\right] \nonumber\\
&= \left( \frac{m_i}{m_i - 2} \right)^2 
\cdot \frac{8 n_i^2 \tilde{\omega}^4 (\nu + 2)^2}{4 (\nu - 2)^2 (\nu - 4)} \nonumber\\[6pt]
&\quad + \frac{2 m_i^2}{k^2 (m_i - 2)^2 (m_i - 4)} 
\left[
\left( k + \frac{n_i \tilde{\omega}^2 (\nu + 2)}{\nu - 2} \right)^2 
+ (m_i - 2) \left( k + \frac{2 n_i \tilde{\omega}^2 (\nu + 2)}{\nu - 2} \right)
\right]
\end{align*}
Hence, 
\begin{align*}
\mathrm{Var}\left( \frac{F_i}{n_i} \right) 
&= \left( \frac{m_i}{m_i - 2} \right)^2 
\cdot \frac{2 \tilde{\omega}^4 (\nu + 2)^2}{(\nu - 2)^2 (\nu - 4)} \\[6pt]
&\quad + \frac{2 m_i^2}{k^2 (m_i - 2)^2 (m_i - 4) n_i^2} 
\left[
\left( k + \frac{n_i \tilde{\omega}^2 (\nu + 2)}{\nu - 2} \right)^2 
+ (m_i - 2) \left( k + \frac{2 n_i \tilde{\omega}^2 (\nu + 2)}{\nu - 2} \right)
\right]
\end{align*}

\noindent
Hence, assuming $m_i = O(n_i)$ and $k = O(1)$ and 
ignoring the $O\left(\frac{1}{n_i}\right)$ and $O\left(\frac{1}{n_i^2}\right)$ terms, the MOM estimator of $\nu$ is:

\begin{equation*}
\hat{\nu}_{\text{MOM}} = 4 + \frac{2 \bar{x}^2}{S^2}
\end{equation*}

where $F_i$s are the observed test statistics, $i = 1,2,\ldots,M$, \[\Bar{x} = \frac{1}{M} \sum_{i=1}^M \left(\frac{F_i}{n_i}\right)\] and \[S^2 = \frac{1}{M}\sum_{i=1}^M \left(\frac{F_i}{n_i} - \frac{1}{M} \sum_{i=1}^M \left(\frac{F_i}{n_i}\right)\right)^2.\]
 \end{document}